\documentclass[lettersize,journal]{IEEEtran}
\usepackage{amsmath,amsfonts}
\usepackage{algorithm}
\usepackage{algorithmicx}
\usepackage{algpseudocode}
\usepackage{array}
\usepackage[caption=false,font=normalsize,labelfont=sf,textfont=sf]{subfig}
\usepackage{textcomp}
\usepackage{stfloats}
\usepackage{url}
\usepackage{verbatim}
\usepackage{graphicx}
\usepackage{cite}
\usepackage{makecell}
\usepackage{amssymb}
\usepackage{xspace}
\usepackage{multirow}

\newtheorem{example}{Example}
\newtheorem{definition}{Definition}

\newcommand{\bi}{\begin{itemize}}
\newcommand{\ei}{\end{itemize}}

\newcommand{\be}{\begin{enumerate}}
\newcommand{\ee}{\end{enumerate}}
\newcommand{\beqn}{\begin{eqnarray*}}
\newcommand{\eeqn}{\end{eqnarray*}}

\newcommand{\stitle}[1]{\vspace{1ex}\noindent{\bf #1}}

\newcommand{\deepdb}{\textit{DeepDB}}
\newcommand{\naru}{\textit{Naru}}
\newcommand{\fctj}{\textit{FactorJoin}}
\newcommand{\uae}{\textit{UAE}}
\newcommand{\mscn}{\textit{MSCN}}
\newcommand{\xgb}{\textit{LW-XGB}}
\newcommand{\nn}{\textit{LW-NN}}
\newcommand{\bayes}{\textit{BayesCard}}
\newcommand{\dmv}{\textit{DMV}}
\newcommand{\covertype}{\textit{Forest}}
\newcommand{\oursys}{\textit{CoDe}\xspace}
\usepackage{xcolor}
\usepackage{color}

\newenvironment{acknowledgment}
{\par\vspace*{12pt}\noindent\textbf{Acknowledgments}\par\noindent}
{}

\begin{document}

\title{A Lightweight Learned Cardinality Estimation Model}

\author{Yaoyu Zhu, Jintao Zhang*, Guoliang Li*, and Jianhua Feng
\thanks{The authors are with the Department of Computer Science and Technology, Tsinghua University, Beijing, China.  E-mail: \{zyy18@mails., zhang-jt24@mails., liguoliang@, fengjh@\}tsinghua.edu.cn
\\ Corresponding
author: Jintao Zhang and Guoliang Li.}
}


\maketitle

\begin{abstract}
Cardinality estimation is a fundamental task in database management systems, aiming to predict query results accurately without executing the queries. However, existing techniques either achieve low estimation accuracy or take high inference latency. Simultaneously achieving high speed and accuracy becomes critical for the cardinality estimation problem. In this paper, we propose a novel data-driven approach called \oursys (Covering with Decompositions) to address this problem. \oursys employs the concept of covering design, which divides the table into multiple smaller, overlapping segments. For each segment, \oursys utilizes tensor decomposition to accurately model its data distribution. Moreover, \oursys introduces innovative algorithms to select the best-fitting distributions for each query, combining them to estimate the final result. By employing multiple models to approximate distributions, \oursys excels in effectively modeling discrete distributions and ensuring computational efficiency. Notably, experimental results show that our method represents a significant advancement in cardinality estimation, achieving state-of-the-art levels of both estimation accuracy and inference efficiency. Across various datasets, \oursys achieves absolute accuracy in estimating more than half of the queries.
\end{abstract}

\begin{IEEEkeywords}
Cardinality estimation, tensor decomposition, covering design.
\end{IEEEkeywords}

\section{Introduction}
\label{sec:introduction}
Cardinality estimation poses a critical challenge in database management systems (DBMS) as it aims to predict query results accurately without executing the queries. This task is crucial for query optimization, as it allows the optimizer to devise the most efficient query plans. Despite numerous proposed solutions, cardinality estimation remains an unsolved problem. Two primary approaches have been explored to tackle this issue: workload-driven methods~\cite{cidr2019/mscn, pvldb/DuttWNKNC19} and data-driven methods~\cite{vldb2020/deepdb, wu2020bayescard, pvldb/YangKLLDCS20}.

\begin{figure}
    \centering
    \includegraphics[width=0.8\columnwidth]{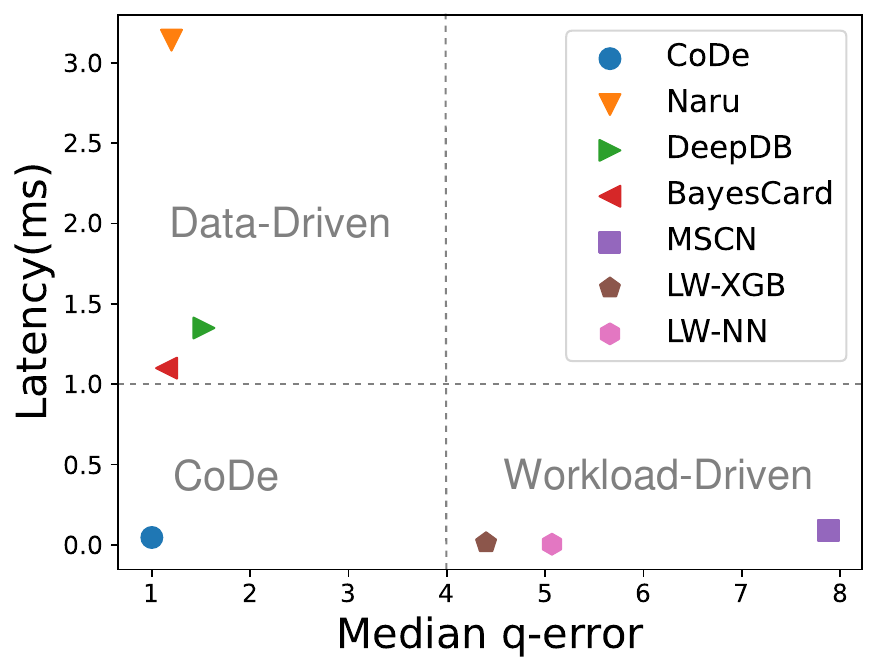}
	\vspace{-1em}
    \caption{Median q-error and Latency on DMV Dataset.}
    \label{fig:compare}
    \vspace{-1em}
\end{figure}

\textbf{Motivation.}
Figure~\ref{fig:compare} illustrates the comparison between our work and the limitations of existing methods. Workload-driven methods focus on learning patterns from historical workloads and their corresponding results. While these methods are generally fast, their accuracy can degrade when workloads change or are randomly generated. This limitation stems from their lack of direct access to the underlying data and their heavy reliance on the distribution of past workloads. As a result, they are positioned in the bottom-right corner of the graph. On the other hand, recent advancements in data-driven methods directly learn the data distribution, significantly improving estimation accuracy. These approaches prioritize accuracy by leveraging the data distribution, but this comes at the cost of slower inference speeds and larger model sizes. Data-driven methods are often orders of magnitude slower than workload-driven methods, placing them in the top-left corner of the graph. Achieving both high speed and accuracy simultaneously is a critical challenge in cardinality estimation, which our work aims to address.

Recent research, such as UAE~\cite{wu2021unified}, has explored hybrid approaches that combine data and workload information, using workload patterns to enhance data learning. However, UAE treats the problem as a two-objective optimization, primarily using workload information to refine data distribution without fully analyzing the workload's inherent patterns. We believe there is significant potential to further analyze workloads, extracting insights that go beyond merely improving data distribution. Such analysis could reveal how queries are formulated, providing valuable information that cannot be derived from the data alone.

In terms of accuracy, it's worth noting that even data-driven methods encounter limitations, particularly in scenarios where the dataset is discrete and the query involves equality in the predicate. We argue that cardinality estimation for discrete and continuous datasets represents two distinct challenges. Discrete values, being nominal and orthogonal, result in queries that only allow equalities in the predicate, evaluating the densities of individual values. In contrast, continuous values, being ordinal, give rise to queries involving inequalities in the predicate, which represent the cumulative densities across a range of values. Consequently, the selection of the query workload becomes a pivotal factor influencing estimation accuracy.
Considering that real-world datasets often exhibit long-tail distributions, the continuous case heavily relies on a small group of high-density values and the values at both ends of the axis, potentially disregarding the majority of low-density values. On the other hand, the discrete case treats every value on the axis equally, making it a more challenging task. Unlike the continuous case, the discrete case necessitates an accurate estimation of all values, rather than just high-density and end-point values. Many existing methods adopt a continuous perspective of the dataset, thus struggling to accommodate discrete queries.

\begin{table}[t!]
    \centering
    {
    \caption{Distribution of $Reg\_Valid\_Date$ in DMV.}
    \label{table:density}
    \begin{tabular}{| c | c | c | c | c |} 
    \cline{1-5}
    Year & 1972-1973 & 1974-2012 & 2013-2016 & 2017-2019 \\
    \cline{1-5}
    Records & 2935 & 195 & 267784 & 11320963 \\ 
    \cline{1-5}
    \end{tabular}
    }
    \vspace{-1em}
\end{table}

\begin{example}
\label{ex:density}
We summarise the distribution of attribute $Reg\_Valid\_Date$ in the DMV dataset as shown in Table~\ref{table:density}. The data analysis reveals a notable trend where the highest number of records is observed between the years 2017 and 2019. In contrast, the period spanning from 1974 to 2012 shows minimal records. Consequently, accurate estimation of range queries, such as determining the number of records after the year 2000, heavily relies on precise density estimations for the years 2017 to 2019. On the other hand, the densities of the years 1974 to 2012 do not significantly impact these range queries and can be considered less important in the estimation process.
\end{example}

To address these challenges, we propose a novel data-driven approach called \textit{CoDe} (\underline{Co}vering with \underline{De}compositions). While the ultimate goal of cardinality estimation is to capture the joint distribution of all attributes, i.e., the global distribution, this distribution is typically too complex to be explicitly modeled. Instead, previous data-driven methods have employed a single model to approximate the global distribution. However, this approach has a limitation: as the domain size of the dataset grows exponentially, the model must either expand proportionally in size or suffer a decline in accuracy. Thus, we propose the covering design technique which uses multiple models, each responsible for modeling a specific local distribution representing a perspective of the global distribution. By doing so, we reduce the dimensionality involved in each model, simplifying the estimation process. 

For each local distribution, we employ the tensor decomposition method for modeling. It is the reason \textit{CoDe} can be fast and accurate. This method involves vector additions and multiplications, which can be efficiently computed in parallel. Consequently, \textit{CoDe} exhibits exceptional speed. Furthermore, to ensure accuracy in the discrete case, we must treat all attribute values equally. Through tensor decomposition, each value is transformed into an entry on a vector. With appropriate training, we can accurately learn and estimate all values within the attribute. The sole constraint of tensor decomposition lies in its sensitivity to the tensor's size. Fortunately, the covering design technique adeptly resolves this issue by diminishing the dimensionality, ensuring \textit{CoDe}'s effectiveness across various scenarios.

Our contributions can be summarized as follows:

(1) Accuracy and Speed: Our method achieves state-of-the-art levels of both accuracy and speed. We have developed an approach that excels in accurately estimating cardinality while also delivering exceptional computational efficiency.

(2) Modeling Discrete Distribution: Our method demonstrates expertise in modeling discrete distributions. Furthermore, we will showcase the capability of our method to effectively model continuous cases as well. This versatility allows us to handle a wide range of distribution types encountered in real-world scenarios.

(3) Novelty: To the best of our knowledge, we are the pioneering researchers to tackle the cardinality estimation problem by employing a combination of the covering design and tensor decomposition method. This innovative approach sets us apart from existing methodologies and opens new avenues for addressing this challenging problem in an entirely unique manner.

\section{Preliminaries}
\label{sec:pre}

\begin{table}[t!]
    \centering
    {
    \caption{Table of Notations.}
    \label{table:notation}
    \begin{tabular}{| c | c |} 
    \cline{1-2}
    $T$, $\mathbb{T}$ & table and its domain \\
    \cline{1-2}
    $T_j$, $\mathbb{T}_j$ & attributes and their domains \\ 
    \cline{1-2}
    $Q$, $\Phi(Q)$ & query and its estimation \\ 
    \cline{1-2}
    $T_Q$ & set of attributes involved in the query \\ 
    \cline{1-2}
    $B_j$ & blocks of the covering design \\ 
    \cline{1-2}
    $k$ & size of the blocks \\ 
    \cline{1-2}
    $t$ & any subset of size $t$ is contained in one block \\ 
    \cline{1-2}
    $\mathcal{A}$, $\mathcal{T}$, $\mathcal{X}$ & tensors \\ 
    \cline{1-2}
    $w$ & weight vector of the decomposition \\ 
    \cline{1-2}
    $R$ & preset rank of the tensor \\ 
    \cline{1-2}
    $A_j$ & decomposition matrices \\ 
    \cline{1-2}
    $a_{i_1,i_2,...,i_m}$ & entries of a tensor \\ 
    \cline{1-2}
    $f_X$ & probability density function defined over $X$ \\ 
    \cline{1-2}
    $N(\times)$, $N(+)$ & number of multiplications and additions \\ 
    \cline{1-2}
    \end{tabular}
    }
    \vspace{-1em}    
\end{table}

In this section, we introduce the framework of the cardinality estimation problem and the two building blocks of our method \textit{CoDe}, i.e., tensor decomposition and covering design. Capturing the distribution of the data is the critical point of cardinality estimation. There are various ways of learning the distribution. We argue that rather than learning itself, accurately reconstructing the data is more important for the problem. Thus we choose tensor decomposition as our learning model. Furthermore, the idea of covering design is to reduce the exploding dimensions of the data, so that the distribution can be captured easier. The table of notations are illustrated in Table~\ref{table:notation}.

\subsection{Cardinality Estimation}
Let $T = \{T_1, T_2, ..., T_n\}$ be a table with $n$ attributes $T_i, 1 \leq i \leq n$, and let $\mathbb{T} = \mathbb{T}_1 \times \mathbb{T}_2 \times ... \times \mathbb{T}_n$ be the domains of the table and its attributes, respectively. There exists a probability density function $f_T : \mathbb{T} \to [0, 1]$. We aim to find the number of records that satisfy the conditions of query $Q$ in $T$, which is denoted as $\Phi(Q)$:
$$
    \Phi(Q) = P(\bigcap_{T_i \in T_Q} T_i = i_j) \cdot |\mathbb{T}| 
    =f_{T_Q}(\cdot) |\mathbb{T}| 
$$
where $T_Q\subseteq T$ is the set of attributes which $Q$ has constraints on, and $f_{T_Q}$ is the local distribution defined on $T_Q$. We omit the expression inside the bracket for convenience purposes, which will be discussed later. The problem of cardinality estimation is equivalent to estimating $f_{T_Q}$.

The multi-table task is considered the same as the single-table scenario, where the distinction lies in the fact that multiple tables are pre-joined into a unified larger table beforehand.

\subsection{Tensor Decomposition}

\begin{figure}
    \centering
    \includegraphics[width=\columnwidth]{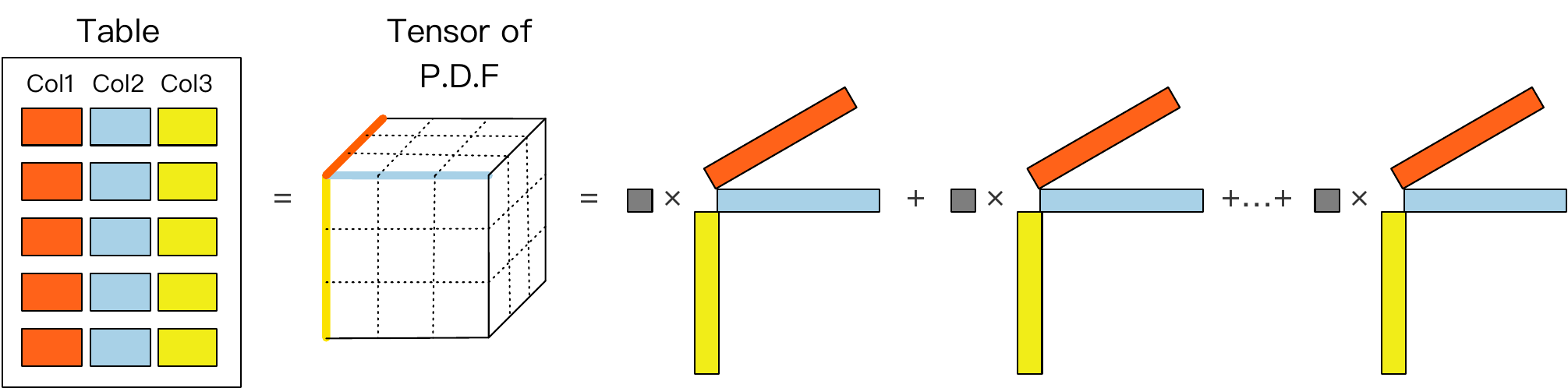}
    \vspace{-1em}
    \caption{Tensor Decomposition.}
    \label{fig:decomp}
    \vspace{-1em}
\end{figure}

In broad terms, tensors can be considered as a generalization and extension of matrices into higher dimensions. Dimensionality reduction plays a vital role in enhancing performance while maintaining expressive capabilities. Tensor decomposition algorithms provide a means of representing tensors using low-dimensional vectors or matrices. Several well-known tensor decomposition techniques include CANDECOMP/PARAFAC (CP) decomposition~\cite{kolda2009tensor}, Tucker decomposition~\cite{tucker1966some}, and tensor train decomposition~\cite{oseledets2011tensor}. For our research, we select CP decomposition due to its simplicity and ease of reconstruction. Furthermore, our method can readily be extended to support other decomposition techniques, offering flexibility and adaptability in future works.

Let $\mathcal{A} \in \mathbb{R}^{d_1 \times d_2 \times ... \times d_m}$ be a $m$-way tensor, then it can be expressed as the linear combination of $R$ rank-$1$ tensors: 
\begin{align}
    \mathcal{A} & \approx [\![ w; A^{(1)}, A^{(2)}, ..., A^{(m)} ]\!] \nonumber \\
    & = \sum_{r=1}^R w[r] \cdot A^{(1)}[:, r] \otimes A^{(2)}[:, r] \otimes ... \otimes A^{(m)}[:, r] \label{eq:decomp}
\end{align}
where $\otimes$ represents the outer product. A tensor is a rank-$1$ tensor if it can be decomposed into the outer product of multiple vectors. 

\begin{example}
\label{ex:decomp}
    A two-way tensor (i.e. a matrix) can be decomposed as follows:
    \begin{align*}
    \begin{bmatrix}
        1 & 2 & 3 \\
        2 & 3 & 4 \\
        3 & 5 & 7
    \end{bmatrix}
    & =
    \begin{bmatrix}
        0 & 1 & 2 \\
        0 & 1 & 2 \\
        0 & 2 & 4
    \end{bmatrix}
    +
    \begin{bmatrix}
        1 & 1 & 1 \\
        2 & 2 & 2 \\
        3 & 3 & 3
    \end{bmatrix} \\
    & = 
    \begin{bmatrix}
        1 \\ 1 \\ 2
    \end{bmatrix}
    \otimes
    \begin{bmatrix}
        0 \\ 1 \\ 2
    \end{bmatrix}
    +
    \begin{bmatrix}
        1 \\ 2 \\ 3
    \end{bmatrix}
    \otimes
    \begin{bmatrix}
        1 \\ 1 \\ 1
    \end{bmatrix}
    \end{align*}
In this example, the rank of the original matrix is $2$, and therefore it can be expressed as the sum of two rank-$1$ matrices.
\end{example}

The rank of a tensor is defined as the smallest number of rank-$1$ tensors required to achieve equality in equation~\eqref{eq:decomp}. Unfortunately, computing the exact rank is NP-hard. Consequently, researchers often determine an empirical rank, denoted as $R$. Additionally, in this paper, the decomposition matrices $A^{(j)}$ are normalized, so that $\sum w = \sum \mathcal{A}$. The rationale behind this normalization will be elaborated upon in Section~\ref{sec:estimation}, where we provide a detailed explanation for this choice.

The CP algorithm suggests that by optimizing the following problem, the decomposition of $\mathcal{X}$ can be seen as the approximation of the decomposition of $\mathcal{A}$: 
\begin{equation}
\label{eq:CP}
    \min_\mathcal{X} ||\mathcal{A}-\mathcal{X}|| \quad s.t. \ \mathcal{X} = \sum_{r=1}^R w[r] \bigotimes_{j=1}^m A^{(j)}[:, r] 
\end{equation}
where $|| \cdot ||$ is the Frobenius norm. Such an optimization problem is often solved by the alternating least squares (ALS) technique, which alternatively updates one of the decomposition matrices $A^{(j)}$ while keeping the others fixed.

With the decomposition, we can compute any entry in the tensor.
\begin{equation}
\label{eq:entry}
    a_{i_1,i_2,...,i_m} = \sum_{r=1}^R \Big( w[r] \prod_{j=1}^m A^{(j)}[i_j, r] \Big)
\end{equation}
$a_{i_1,i_2,...,i_m}$ indicates the $(i_1,i_2,...,i_m)$th entry in the tensor.

\subsection{Covering Design}
The domain size of the dataset grows exponentially as the number of attributes increases. This poses a challenge for the tensor decomposition algorithm, which may encounter difficulties when dealing with excessively wide tables. In such cases, it becomes crucial for the covering design to come into play and provide a solution. The covering design, as a mathematical concept, plays a pivotal role in our approach by facilitating the selection of multiple subsets of the table $T$ in a strategically meaningful manner. This enables us to effectively decompose the original wide table $T$ into multiple narrower tables, each of which represents a distinct perspective of the overall data. It is formally defined as follows:
\begin{definition}
Given a finite set $V=\{1,2,...,v\}$, \emph{blocks} are a collection of $k$-element subsets of $V$. A $C(v,k,t)$ \emph{covering design} is a set of blocks such that any $t$-element subset ($t < k$) is contained in at least one block.
\end{definition}

\begin{example}
\label{ex:cover}
    Let $v=7, k=4, t=2$, then $C(7,4,2)$ is a set of 4-element subsets of $V = \{1,2,...,7\}$ that covers all 2-elements subsets of $V$. A possible set of blocks of $C(7,4,2)$ are:
    $$
        \{1,2,3,4\},\{1,4,5,6\},\{1,5,6,7\},\{2,3,4,7\},\{2,3,5,6\}
    $$
\end{example}

There is a vast amount of work studying the lower bound of covering designs and corresponding blocks~\cite{gordon1995new, colbourn2010crc, dinitz1992contemporary}. Many of the best solutions are collected by Dan Gordon~\cite{misc_dan_gordon}.

\section{\oursys Overview}
\label{sec:overview}

\subsection{High Level Idea}
Rather than capturing the entire distribution within a singular expansive model, our approach employs a multi-model strategy to collectively encompass the joint distribution. Each smaller-scale model is tasked with representing the discrete probability density function defined on a subset of attributes, achieved through the tensor decomposition technique. The selection of these attribute subsets is determined by the covering design methodology. 

\subsection{Pipeline of Training and Estimation.} 

\begin{figure*}[t!]
    \centering
	\includegraphics[width=\textwidth]{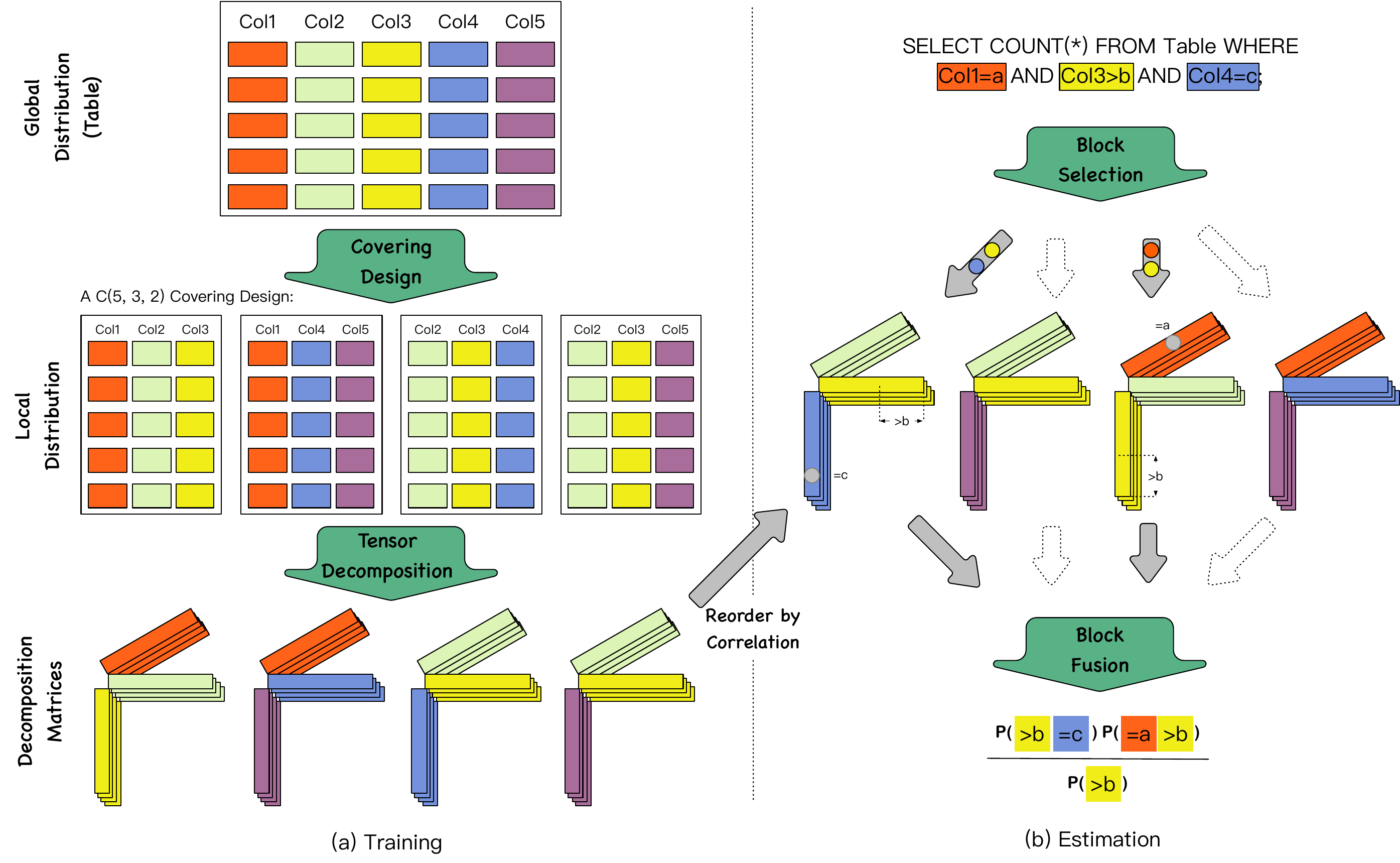}
    \vspace{-1em}
    \caption{An Example of the Workflow of \oursys. (a) Offline Training Process, (b) Online Estimation Process.}
    \label{fig:example}
    \vspace{-1em}
\end{figure*}

For offline training, unlike traditional approaches that rely on a single large-scale model to learn the distribution, the \textit{CoDe} method adopts the covering design that incorporates multiple distinct local distributions to capture diverse perspectives of the global distribution. The table is divided into multiple smaller overlapping tables, each comprising $k$ attributes representing a local distribution of $T$. The covering design methodology ensures that any query involving no more than $t$ filters (where $t<k$) can be resolved by a single local distribution. Since the table is discretized, the probability density function (P.D.F) of a local distribution takes the form of a $k$-dimensional tensor. These local distributions are organized based on attribute correlation, prioritizing closely related local distributions for selection during estimation. The \textit{CoDe} method decomposes these P.D.F tensors into decomposition matrices using the CP algorithm, with the results stored for online processing. All decomposition matrices are normalized with L1 distance for the convenience of calculation. Section~\ref{sec:train} introduces the model training.

For online cardinality estimation, given a specific query, \textit{CoDe} first identifies the most relevant local distributions that align with the query. When the number of filters is sufficiently small, a single local distribution is often adequate. The tensor decomposition method offers a straightforward approach for computing individual tensor entries. Query estimation is achieved by summing a sequence of tensor entries. However, if a single local distribution is insufficient to cover the query adequately, multiple local distributions can be selected simultaneously. In this case, attributes that are not covered by the same local distribution are assumed to be independent. \textit{CoDe} optimally minimizes the number of selected local distributions and the associated independence assumptions. Probabilities for each local distribution are calculated individually and then combined using Bayes' theorem to produce a comprehensive estimation. Notably, \textit{CoDe} showcases its proficiency in handling continuous datasets as well. Section~\ref{sec:estimation} introduces query estimation.

\subsection{Necessity of Covering Design and Tensor Decomposition} 

Covering design is essential because, like conventional partitioning methods, it reduces dimensionality—a critical step for subsequent processing. However, covering design possesses unique capabilities that partitioning methods lack. Traditional partitioning divides attributes into non-overlapping or minimally overlapping blocks, which can result in queries involving attributes scattered across multiple blocks. For optimal performance, all attributes in a query should ideally be addressed by a single model; otherwise, determining their joint probability requires impractical independence assumptions. This limitation explains the inaccuracies observed in prior approaches.
In contrast, covering design guarantees that any combination of up to $t$ attributes can be resolved within a single block, a condition that encompasses most real-world queries. This structural advantage is the foundation of CoDe’s exceptional performance, enabling an unprecedented $50\%$ absolute improvement in accuracy compared to existing methods.

While tensor decomposition may not be the most accurate method in isolation, its integration with dimensionality reduction via covering design enables substantial improvements in accuracy across various methods, including tensor decomposition itself. Beyond accuracy, computational efficiency is a critical consideration for our task, and tensor decomposition excels in this regard. Although it demands significant offline training time, it achieves exceptionally fast online inference—a trade-off whose complexity we rigorously analyze in Section~\ref{sec:estimation}.
The synergy between covering design and tensor decomposition is pivotal. For high-dimensional datasets, tensor decomposition alone faces scalability challenges due to its training complexity, which grows with the domain size of the data. However, when preceded by dimensionality reduction through covering design, tensor decomposition becomes both computationally tractable and highly accurate. This combination thus offers an optimized solution, balancing speed and precision effectively.


\section{Model Training}
\label{sec:train}

\subsection{Tensor Decomposition}

\textit{CoDe} operates by representing the distribution of the table $T$ as a high-dimensional tensor denoted as $\mathcal{T}$. For categorical attributes, each unique value is treated as an entry on the tensor axis. However, for numerical or ordinal attributes with a large number of distinct values, we discretize the data into appropriate intervals, assigning each interval a unique entry on the tensor axis.

Ideally, there should exist a natural decomposition of $\mathcal{T}$. However, finding such a decomposition is incredibly challenging due to the potentially high dimensionality and immense rank of the joint distribution.

Let's first address the rank problem. Notably, $\mathcal{T}$ comprises integer entries since the number of records must be a whole number. The CP algorithm, along with Equation~\eqref{eq:CP}, approximates the decomposition by performing the decomposition on a tensor $\mathcal{X}$ that is sufficiently close to $\mathcal{T}$. More precisely, we can find a tensor $\mathcal{X} \in \mathbb{R}^n$ satisfying $\max(|\mathcal{T} - \mathcal{X}|) < 0.5$, where $|\cdot|$ denotes absolute values. Any real entry in $\mathcal{X}$ will be rounded to the integer represented by $\mathcal{T}$. The rank of $\mathcal{X}$ can be considerably smaller than that of $\mathcal{T}$. Consequently, by selecting a relatively small value for $R$, $\mathcal{X}$ can be rounded to approximate $\mathcal{T}$.

\begin{example}
    Consider the matrix provided in example~\ref{ex:decomp}, which consists of integer values and has a rank of 2. Remarkably, it is possible to find a real matrix that closely approximates the original matrix while possessing a lower rank of 1.
    $$
    \begin{bmatrix}
        1 & 2 & 3 \\
        2 & 3 & 4 \\
        3 & 5 & 7
    \end{bmatrix}
    \approx
    \begin{bmatrix}
        1 & 1.8 & 2.6 \\
        1.6 & 2.88 & 4.16 \\
        2.8 & 5.04 & 7.28
    \end{bmatrix}
    =
    \begin{bmatrix}
        1 \\ 1.8 \\ 2.6
    \end{bmatrix}
    \otimes
    \begin{bmatrix}
        1 \\ 1.6 \\ 2.8
    \end{bmatrix}
    $$
\end{example}

Next, let's address the challenge of high-dimensional joint distributions. We define the domain size, denoted as $\eta(S)$, for a subset $S \subseteq T$ as follows:
$$
    \eta(S) = \prod_{T_i \in S} |\mathbb{T}_i|
$$
We assume that the difficulty of decomposing a tensor over $S$ is proportional to its domain size $\eta(S)$. Hence, there exists an upper bound, denoted as $M_k$, beyond which the decomposition process fails due to the excessive dimensions, i.e., $\eta(S) \leq M_k$. In practice, $M_k$ is often determined by the memory of the device.

\subsection{Covering Design}

The graph depicted in Figure~\ref{fig:example}(a) highlights the significance of the covering design concept in simplifying the intricate distribution of the table $T$. Rather than attempting to learn the complex joint distribution of $T$, we adopt an approach where the attributes of $T$ are distributed across multiple blocks. 
To learn the local distribution specific to each block, we aggregate the dataset according to the block to acquire the P.D.F tensor of that block. We then use the CP algorithm to decompose these tensors.
This methodology enables us to effectively capture the distribution of $T$ from diverse perspectives. One notable advantage of this approach is that any query involving constraints on no more than $t$ attributes can be accommodated within a single block. 

\begin{example}
Figure~\ref{fig:sampling} visually demonstrates the relationship between the number of filters in the predicate and the possibility of a zero query result in the DMV dataset. The findings indicate that when the number of filters exceeds $5$, the likelihood of a randomly generated query returning a non-zero value diminishes significantly, with a high probability of yielding a result of $0$. In the upcoming experiment section, we will provide empirical evidence supporting the notion that estimating a zero-query with \textit{CoDe} is relatively straightforward. Consequently, our primary objective is to accurately estimate queries with fewer filters.

Fortunately, the implementation of the covering design technique assures us that any subset of size $t$ will be covered by at least one block. This important guarantee implies that queries involving a smaller number of filters are more likely to be covered by a single block. This simplifies the estimation process, as the relevant information necessary for accurate estimation can be contained within a single block.

\begin{figure}
    \centering
    \includegraphics[width=\columnwidth]{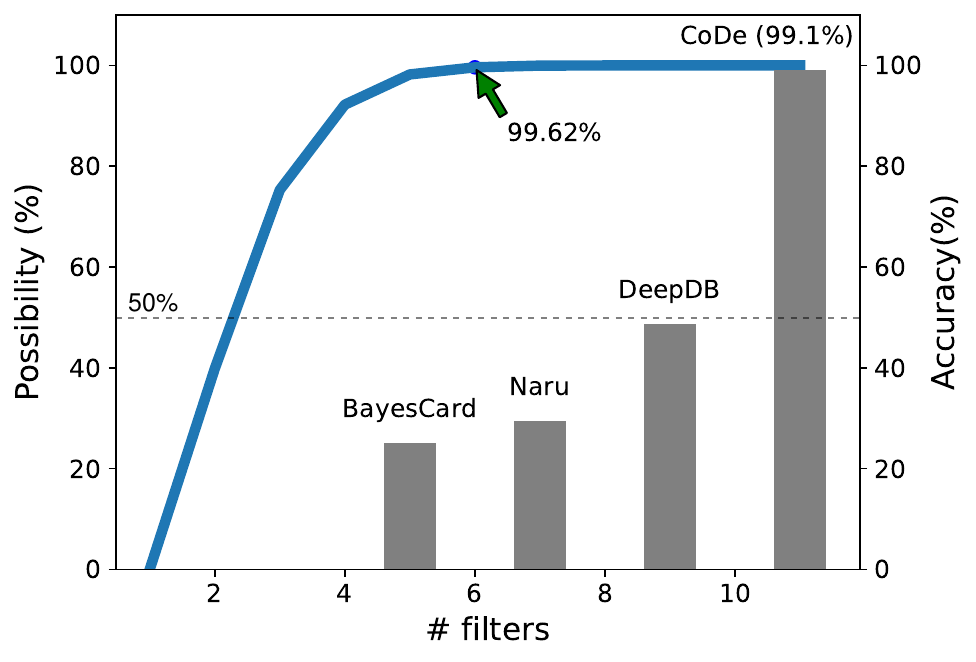}
	\vspace{-1em}
    \caption{Possibility Queries Returns Zero Answer (Curve), and Methods Accuracy on Zero-queries (Histogram).}
    \label{fig:sampling}
    \vspace{-1em}
\end{figure}

\end{example}

To optimize efficiency, our goal is to maximize the coverage of subsets while minimizing the number of blocks required. This means we aim for large values of both $k$ and $t$. When the domain size of a block is sufficiently small, we can compute its decomposition. Therefore, the selection of a covering design is constrained by the parameter $M_k$. In other words, we strive to find the highest value of $k$ that satisfies the condition $\eta(B_i) \leq M_k$ for all blocks $B_i$ in the set $C(n, k, t)$. The choice of $t$ is determined empirically, allowing flexibility in its selection, as long as it ensures the number of blocks does not become excessively large.

Building upon our previous approach, we take a significant leap forward in addressing the challenges posed by attributes with varying domain sizes. We have observed that attributes exhibit substantial differences in their domain, ranging from only two distinct values to several hundred. Consequently, the domain sizes of blocks may vary significantly, potentially differing by orders of magnitude.

Considering that the domain size directly influences the training difficulty, we recognize that the choice of $k$ is primarily limited by the block with the largest domain size. In our algorithm, we also acknowledge that the blocks do not need to be of equal size, allowing for variations in $k$ across different blocks. This realization leads us to reframe the problem statement: Can we identify a set of blocks ${B_i}$ with different sizes, yet similar domain sizes, satisfying $\eta (B_i) \leq M_k \ \forall i$? Furthermore, can we ensure that every subset with a domain size smaller or equal to $M_t$ is contained within at least one block? Here, $M_t$ represents a generalization of the constant $t$, which defines the domain size of the subsets to be covered.

However, it is important to note that this problem is NP-hard. As a result, we propose a heuristic solution to tackle it effectively. Our approach involves combining multiple inferior attributes through the Cartesian product, creating a new superior attribute. Subsequently, we apply a classic covering design strategy to this new set of attributes. By employing this method, we can achieve a reduction in the number of blocks required, accompanied by smaller $k$ values, thus optimizing the overall efficiency of our approach.

\begin{table}[t!]
    \centering
    {
    \caption{An Example of Domain Sizes.}
    \label{table:cover1}
    \begin{tabular}{| c | c | c | c | c | c | c | c |} 
    \cline{1-8}
    attributes & 1 & 2 & 3 & 4 & 5 & 6 & 7 \\
    \cline{1-8}
    domain sizes & 10 & 11 & 3 & 4 & 7 & 2 & 2\\ 
    \cline{1-8}
    \end{tabular}
    }
    \vspace{-1em}    
\end{table}
\begin{table}[t!]
    \centering
    {
    \caption{An Example Domain Sizes of Joint Attributes.}
    \label{table:cover2}
    \begin{tabular}{| c | c | c | c | c | c |} 
    \cline{1-6}
    attributes & 1 & 2 & 3 $\times$ 6 & 4 $\times$ 7 & 5 \\
    \cline{1-6}
    domain sizes & 10 & 11 & 6 & 8 & 7\\ 
    \cline{1-6}
    \end{tabular}
    }
    \vspace{-1em}    
\end{table}

\begin{example}
\label{ex:join}
    Consider a table with $7$ attributes that have domain size shown in table~\ref{table:cover1}, it has a covering design shown in the example~\ref{ex:cover}.
    We join attribute $3$ with $6$, and $4$ with $7$, then we have domain sizes shown in table~\ref{table:cover2}, and apply a $C(5, 3, 2)$ covering design on it, giving $4$ blocks as follows:
    $$
        \{1,2,3,6\}, \{1,4,5,7\}, \{2,3,4,6,7\}, \{2,3,5,6\}
    $$
    After the process, we have one block less than the previous covering.
\end{example}

We summarize the procedure for the covering design as follows:
(1) Combining Attributes (if necessary): This step involves merging attributes, as illustrated in Example~\ref{ex:join}. Whether or not this step is applied, the parameter $v$ remains fixed.
(2) Choosing $k$: The value of $k$ is generally preferred to be as large as possible. However, it is constrained by memory capacity. During training, the size of a tensor block equals the product of the number of distinct values across the $k$ selected attributes. We recommend selecting the largest possible $k$ that allows the tensor to fit within the available memory.
(3) Choosing $t$: A larger $t$ results in more blocks being trained. Users can select $t$ based on a manageable number of blocks for their specific requirements.
(4) Extraction and Training: Given $C(v,k,t)$, we can search for the covering design using existing solutions~\cite{misc_dan_gordon}. Subsequently, for each block, we extract the relevant set of attributes from the dataset and train a minor model based on it.

\subsection{Updates}

When updates occur in the dataset, the approach to updating the model depends on the extent of the changes. Two distinct methods can be employed.

For minor updates, where the dataset undergoes relatively small changes, we can assume that the distribution of the dataset remains constant after updating. Therefore, it suffices to only update the weights, denoted as $w$, for each model. The values of $w$ are adjusted proportionally to maintain the equality $\sum w = \sum \mathcal{T}$, ensuring consistency with the updated dataset. i.e.,
\begin{equation}
    w_{new} = w_{old} \cdot \frac{\# present \ total \ records}{\# previous \ total \ records}
\end{equation}
This approach allows us to incorporate the changes efficiently without the need for a complete model retraining process.

In the case of major updates, where more significant changes are introduced to the dataset, it becomes necessary to update not only the weights but also the decomposition matrices. Leveraging the capabilities of the CP algorithm, we can continue training the model using the previous results as a starting point after updating the weight. It's important to note that while this approach may allow for time savings compared to complete model retraining, achieving optimal results might still require considerable iterations due to the algorithm's convergence properties.

\section{Query Estimation}
\label{sec:estimation}

In this section,  our focus centers on the algorithms utilized in the cardinality estimation procedure. First, we present the computations required within each individual block, subsequently delving into the process of merging multiple blocks. These processes give rise to two crucial criteria that guide the construction of our block selection algorithm. Finally, we discuss the scenarios involving range queries on continuous attributes. 

\subsection{Reconstruct Tensor}

We start with the case when there is only one block, i.e., $k = n$. We have a decomposition of the tensor representing the entire table $T$, i.e., $\mathcal{T} = f_T \cdot |\mathbb{T}| = [\![ w; A^{(1)}, A^{(2)}, ..., A^{(m)} ]\!]$. The density of $Q$ is the sum of all entries that are restricted by $Q$. To be precise, it is an integral (or sum) of entries along the dimensions $T \setminus T_Q$. Recall that the density of an entry is given by equation~\eqref{eq:entry}, the estimation of query $Q$, which is denoted as $\Phi(Q)$, is the sum of entries $a_{i_1, i_2,...,i_n}$ where $i_j$ is either determined by the query or it is a random variable that we will sum up all the distinct values. Hence $\Phi(Q)$ is expressed as follows: 
\begin{align}
\begin{split}
    \Phi(Q) & =\sum_{\forall i_j \in \mathbb{T}_j, T_j \in T \setminus T_Q} a_{i_1, i_2,...,i_n} \\
    & = \sum_{\forall i_j \in \mathbb{T}_j, T_j \in T \setminus T_Q} \sum_{r=1}^R \Big( w[r] \prod_{h=1}^n A^{(h)}[i_h, r] \Big) \\
    & = \sum_{r=1}^R w[r] \prod_{T_h \in T_Q} A^{(h)}[i_h,r] \prod_{T_j \in T \setminus T_Q}  \sum_{i_j \in \mathbb{T}_j} A^{(j)}[i_j,r] 
\end{split}
\end{align}

It is crucial that the decomposition should be normalized with L1-distance, i.e., $\sum_{i_j \in \mathbb{T}_j} A^{(j)}[i_j, r] = 1 \ \forall j,r$. Hence a massive number of calculations can be omitted, and we have:
\begin{equation}
\label{eq:phi}
    \Phi(Q) = \sum_{r=1}^R w[r] \prod_{T_h \in T_Q} A^{(h)}[i_h,r]
\end{equation}
We can see that the estimation of $\Phi(Q)$ only consists of multiplications and additions, which all can be computed in parallel. Furthermore, the number of multiplications $N(\times)$ and additions $N(+)$ are given as follows:
\begin{align}
\label{eq:numcal}
    \begin{split}
        N(\times) & = |T_Q| \cdot R \\
        N(+) & = R - 1 
    \end{split}
\end{align}

$w[r] \prod_{T_h \in T_Q} A^{(h)}[i_h,r]$ requires $|T_Q|$ multiplications and no addition ($w[r]$ also need one multiplication). For $R$ iterations and summing up, it gives the result as in equations~\ref{eq:numcal}.

Consequently, our algorithm exhibits a notable advantage in terms of computational efficiency, with a total number of calculations generally lower than many other existing algorithms. Notably, the algorithm entails significantly more multiplications than additions. Both multiplications and additions scale in proportion to the rank $R$. The number of multiplications is also influenced by the number of filters in the query.

\subsection{Fusion of Blocks}

The covering design technique proves highly effective when $|T_Q| \leq t$. However, The algorithm fails to provide a complete solution when $T_Q$ cannot be covered by any single block. Therefore we propose to use multiple blocks to estimate $\Phi(Q)$.

We start from the case that two blocks $B_1$ and $B_2$ cover the $T_Q$. 
Let the set of common attributes of $B_1$ and $B_2$ be $S_0$. Thus there exist mutually disjoint subsets $S_1$, $S_2$ and $S_0$, such that $B_1 = S_1 \cup S_0$ and $B_2 = S_2 \cup S_0$. We have the probability density functions $f_{B_1} = f_{S_1,S_0}$ and $f_{B_2} = f_{S_2,S_0}$. The probability density function $f_{S_0}$ can be derived from any of the two functions $f_{B_1}$ and $f_{B_2}$ as it is the marginal distribution of the two. We estimate the joint distribution over $B_1 \cup B_2$ as follows:
\begin{equation}
\label{eq:fusion}
    f_{B_1 \cup B_2} \approx \frac{f_{S_1,S_0} \cdot f_{S_2,S_0}}{f_{S_0}}
\end{equation}
The formula relies on the conditional independence assumption: $S_1$ and $S_2$ are independent given $S_0$. Mathematically, this means:$f_{S_1,S_2|S_0} \approx f_{S_1|S_0} \cdot f_{S_2|S_0}$. Therefore, by Bayes Theorem, we have:
\begin{align}
\begin{split}
    f_{B_1 \cup B_2} = f_{S_1,S_2,S_0} & = f_{S_1,S_2|S_0} \cdot f_{S_0} \\
    & \approx f_{S_1|S_0} \cdot f_{S_2|S_0} \cdot f_{S_0} \\
    & = \frac{f_{S_1,S_0}}{f_{S_0}} \cdot \frac{f_{S_2,S_0}}{f_{S_0}} \cdot f_{S_0} \\
    & = \frac{f_{S_1,S_0} \cdot f_{S_2,S_0}}{f_{S_0}}
\end{split}
\end{align}

In practice, we must make an assumption that all attributes in $S_1$ are independent of attributes in $S_2$ under the condition $S_0$. There are in total $|S_1| \cdot |S_2|$ such independence to be made.

In general, a density function requiring multiple blocks is estimated as follows:
\begin{equation}
\label{eq:multi_fusion}
    f_{T_Q} \approx \prod_{B_j s.t. T_Q \subseteq \cup B_j} \frac{f_{B_j \cap T_Q}}{f_{\bigcup_{h<j} B_h \cap B_j \cap T_Q}}
\end{equation}
for the first block $B_j$ covering $T_Q$, the denominator $f_{\bigcup_{h<j} B_h \cap B_j \cap T_Q}$ is defined to be $1$, in particular, $\bigcup_{h<j} B_h$ is considered as an empty set.
The formula iteratively constructs $f_{T_Q}$. It starts with the first block $B_1$, contributing $f_{B_1 \cap T_Q}$. For each subsequent block $B_j$, the result is multiplied by the conditional probability deduced from $B_j$. In detail, the numerator is the probability of attributes covered by $B_j$ in $T_Q$, and the denominator is the probability of overlapping attributes which also covered by previous blocks. The multiplication can be made because of the assumption of conditional independence illustrated in equation~\eqref{eq:fusion}.

Finally, we analyze the computational complexity of Equation~\ref{eq:multi_fusion}. The equation involves only multiplications, additions, and a few divisions, and their upper bounds are calculated as follows:
\begin{align}
    \begin{split}
        N(\times) & < 2 \cdot N(B) \cdot |T_Q| \cdot R + N(B) - 1 \\
        N(+) & = 2 \cdot N(B) \cdot(R - 1) \\
        N(\div) & = N(B)
    \end{split}
\end{align}
where $N(B)$ is the minimum number of blocks required to cover $T_Q$.
As shown in Equations~\ref{eq:numcal}, each probability density function requires fewer than $|T_Q| \cdot R$ multiplications and $R-1$ additions. With $N(B)$ iterations, each connected by multiplications, an additional $N(B) - 1$ multiplications are introduced. Each iteration involves two probability density functions and one division. It is important to note that this upper bound for multiplications is relatively loose, as most probability density functions typically consider far fewer blocks than $T_Q$.

\subsection{Selection of Blocks}

Queries have the potential to involve numerous combinations of attributes. Therefore it is essential to select proper blocks for different queries. Our primary concern is to minimize errors arising from the tensor decomposition and fusion of these selected blocks. The error resulting from tensor decomposition is inevitable and thus can be neglected, leaving us to focus on the error caused by the fusion of blocks.
To address these concerns, we propose two essential criteria for our algorithm:
\be
    \item First, the algorithm must minimize the number of independence assumptions made.
    \item Second, the algorithm should select independence assumptions that minimize errors.
\ee

However, simultaneously satisfying both criteria presents a challenging optimization problem. To tackle this complexity, we propose a greedy algorithm as a practical solution.
First, let's focus on the second criterion of selecting the independence assumptions with minimal errors. In this context, we achieve this goal by breaking the link only between attributes exhibiting the lowest correlation, thereby giving preference to blocks consisting of highly correlated attributes.

One way of measuring the correlation is to evaluate the marginal error resulting from the assumption of independence between two attributes. From the equation~\eqref{eq:fusion}, the marginal error is defined to be:
\begin{align}
    e & = q(f_{T_j,T_h|S_0}, f_{T_j|S_0} \cdot f_{T_h|S_0}) \nonumber \\
    & \approx q(f_{T_j,T_h}, f_{T_j} \cdot f_{T_h}) \nonumber \\
    & \approx \sum_{\mathbb{T}_j, \mathbb{T}_h} |f_{T_j,T_h} - f_{T_j} \cdot f_{T_h}| \ / \sum_{\mathbb{T}_j, \mathbb{T}_h} f_{T_j,T_h} \nonumber \\
    & = \sum_{\mathbb{T}_j, \mathbb{T}_h} |f_{T_j,T_h} - f_{T_j} \cdot f_{T_h}| \label{eq:error}
\end{align}
We first assume the correlation between two attributes remains the same under any condition $S_0$. Next, computing the relative error of an entry in $f_{T_j, T_h}$ can be complicated. Instead, we consider the sum of the absolute error divided by the sum of $f_{T_j, T_h}$. This ratio enables us to estimate the average relative error effectively. 
With the marginal error, we are able to evaluate the cost of the independence assumption, and hence rank the importance of the blocks. 

\begin{algorithm}
\caption{Ordering Blocks by Importance}
\label{algo:order}
\begin{algorithmic}[1]
	\State \textbf{Input:} $f_T$: the table with its records; 
    \State $\{B_l\}$: unordered list of blocks\; 
	\State \textbf{Output:} $\{B_l\}$: reordered blocks\;
    \For{each $T_j,T_h \in T$}
        \State $cor_{j,h} \gets \sum_{\mathbb{T}_j, \mathbb{T}_h} |f_{T_j,T_h} - f_{T_j} \cdot f_{T_h}|$\;
        \State $cor_{j,j} \gets 1$\;
	\EndFor
    \For{each $B_l \in \{B_l\}$}
        \State $r_l \gets \frac{1}{|B_l|^2} \sum_{T_j,T_h \in B_l} cor_{j,h}$\;
	\EndFor
    \State $\{B_l\} \gets$ reorder $\{B_l\}$ based on $r_l$\;
	\State \textbf{return} {$\{B_l\}$}:\
\end{algorithmic}
\end{algorithm}

\stitle{Algorithm~\ref{algo:order}: Ordering Blocks by Importance.}
The algorithm takes as input the local distributions of the table and the list of blocks, and it produces an output comprising reordered blocks ranked by importance. Initially, it computes the marginal error between every pair of attributes using Equation~\eqref{eq:error} (lines~3--6). Subsequently, for each block, the average marginal error of all possible pairs of attributes is calculated (lines~7--9). This calculation reveals how attributes are correlated within each block, serving as an indicator of their relative importance, which directly addresses the second criterion. By reordering the blocks based on these importance scores, the blocks at the top of the reordered list gain higher priority for selection (line~10).
It is important to note that this entire process is independent of the query, making it feasible to perform offline during the training process.

On the other hand, in order to reduce the number of independence assumptions, it is necessary to efficiently cover $T_Q$ with blocks, ensuring the densest coverage using the fewest blocks possible.

\begin{example}
    Let us adopt the covering design stated in example~\ref{ex:cover}. Consider a query involving attributes $\{1,3,5,7\}$. The query requires at least two blocks to be covered. One way is to cover the query with blocks $\{1,4,5,6\}$ and $\{2,3,4,7\}$. By doing so, we assume that attributes $\{1,5\}$ are independent of attributes $\{3,7\}$, which yields $4$ independence assumptions. The other way is to cover the query with blocks $\{1,2,3,4\}$ and $\{1,5,6,7\}$. By doing so, we assume that attribute $\{3\}$ is independent of attributes $\{5,7\}$ under the condition $\{1\}$, which yields $2$ independence assumptions. We conclude that the latter is the better solution.
\end{example}

\begin{algorithm}
\caption{Selection of Blocks}
\label{algo:select}
\begin{algorithmic}[1]
	\State \textbf{Input:} $\{B_l\}$: list of blocks with order; 
    \State $T_Q$: the set of attributes that the query has constraints on\;
	\State \textbf{Output:} $C$: selected covering blocks\;
	\State \textbf{Initialize}: $\{score_j\} = 0$; $C = \emptyset $; $remains = T_Q$; $\alpha = 0.01$\;
    \While{$|remains|>0$}
        \State $score \gets score \cdot \alpha$\;
        \For{each $B_l \in \{B_l\}$}
            \State $score_l \gets score_l + |B_l \cap remains|$\;
        \EndFor
        \State $high \gets$ index of upmost $score$\;
        \State $C \gets C \cup B_{high}$\;
        \State $remains \gets remains \setminus B_{high}$\;
    \EndWhile
	\State \textbf{return}{$C$}:\
\end{algorithmic}
\end{algorithm}

\stitle{Algorithm~\ref{algo:select}: Selection of Blocks.}
The algorithm takes the query as input and generates a set of blocks that efficiently cover $T_Q$. It begins by initializing the scores for each block to be $0$, an empty set $C$ to store selected blocks, a set $remains$ to record the uncovered attributes in $T_Q$, and a decay rate $\alpha$ set to $0.01$.
The algorithm iterates continuously as long as $remains$ is not empty (lines~4--12). During each iteration, it counts the number of common attributes between each block and the attributes in $remains$. For each common attribute, the corresponding block's score is incremented by $1$ (lines~6--8). The algorithm then selects the block with the highest score. If multiple blocks have the same score, the algorithm selects the one with the highest priority from the list, adhering to the second criterion (lines~9--10). The selected block's attributes are removed from $remains$ (line~11).
Before proceeding to the next iteration, the algorithm decays the scores of all blocks by a factor of $\alpha$ (line~5). This ensures that attributes covered in the current round are given higher importance than those covered in previous rounds. The reason for this decay process is to encourage the selection of blocks that not only cover uncovered attributes but also overlap with each other as much as possible. However, emphasizing the coverage of uncovered attributes takes precedence due to this decay mechanism.
Importantly, all of these processes are calculated online.

The two criteria are not mutually exclusive, but we prioritize the first one (minimizing independence assumptions) as it is considered more critical. Specifically, only when two options share the same number of independence assumptions do we select the one that produces fewer errors. The offline reordering algorithm ranks blocks based on the internal correlation of their attributes. Blocks with highly correlated attributes are deemed more significant and are thus prioritized. For online selection, a greedy algorithm is employed to minimize the number of blocks required to cover $T_Q$, while simultaneously maximizing overlaps in shared attributes. This approach inherently reduces the number of independence assumptions needed for computation.

\subsection{Continuous Case}

\begin{figure}
    \includegraphics[width=\columnwidth]{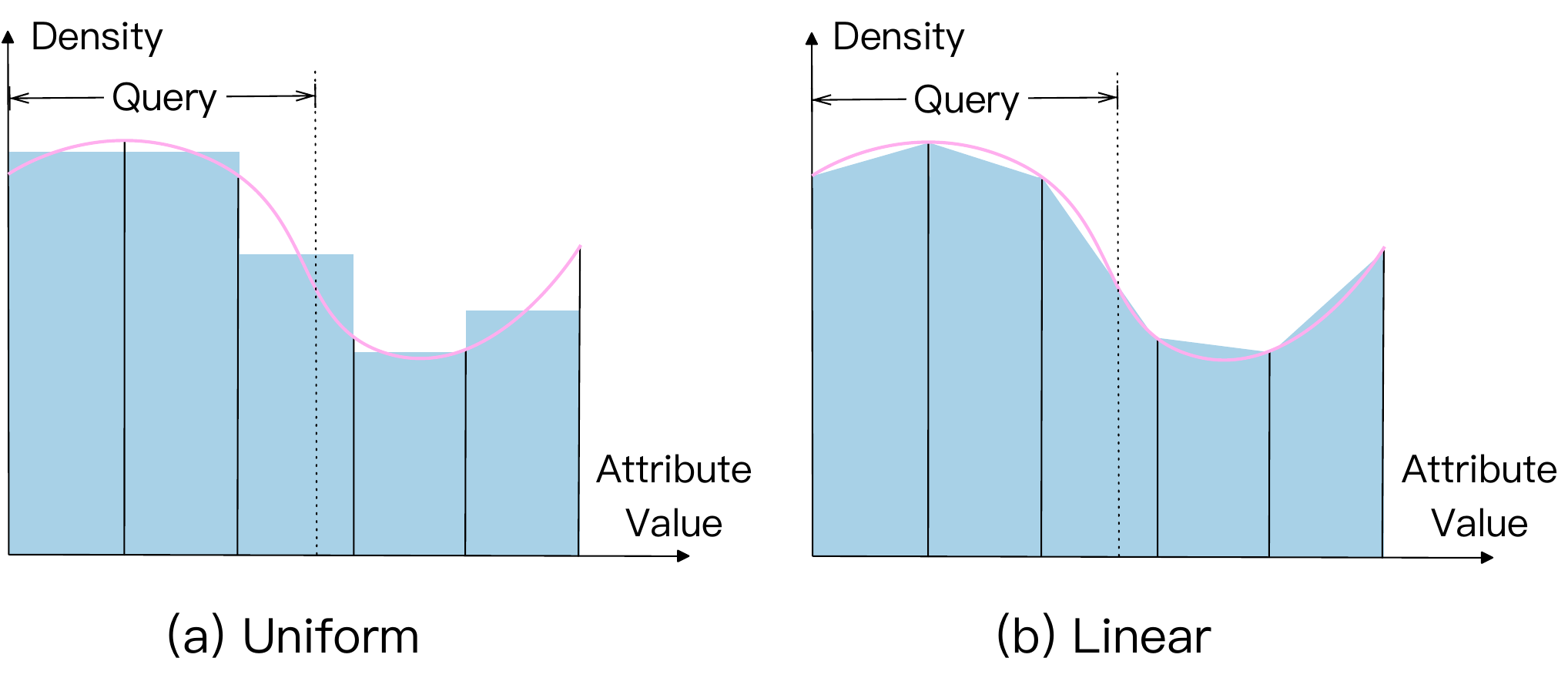}
	\vspace{-1.5em}
    \caption{Data Distributions on Continuous Intervals.}
    \label{fig:ctn}
    \vspace{-1em}
\end{figure}

A range query involving continuous attributes can be conceptually considered as the integration or summation of queries on discrete points. However, such calculations can be significantly slower compared to point queries. To overcome this challenge, the tensor decomposition strategy offers an alternative approach. Leveraging the multiplicative distribution law, we can first compute the marginal probability of each rank-$1$ tensor (distribution), followed by multiplications, and then sum up the results from these rank-$1$ tensors.
Given that the attributes are discretized, the marginal probability, restricted by the query, can be viewed as a linear combination of densities over intervals, with $\xi_j$ representing their respective coefficients. In general, the query estimation is expressed as:
\begin{equation}
\label{eq:phicon}
    \Phi(Q) = \sum_{r=1}^R w[r] \prod_{T_h \in T_Q^{(D)}} A^{(h)}[i_h,r] \prod_{T_j \in T_Q^{(C)}}  \sum_{i_j \in \mathbb{T}_j} \xi_j A^{(j)}[i_j,r] 
\end{equation}

where $T_Q^{(D)}$ and $T_Q^{(C)}$ refer to the sets of discrete attributes and continuous attributes, respectively, which the query has constraints on.
This equation differs from Equation~\ref{eq:phi} because, for continuous attributes, it sums the densities of selected intervals, each weighted by different coefficients, rather than selecting a single categorical value.
When the query fully covers an interval, the corresponding coefficient is set to $1$. However, for intervals that are only partially covered by the query, their coefficients need to be estimated. 

This is not a standard interpolation problem, as interpolation typically relies on known points to estimate the points in between. In our tensor, entry values represent interval integrals rather than specific points, rendering the direct application of linear or cubic interpolation inappropriate.
As illustrated in Figure~\ref{fig:ctn}, the distribution within each interval is unknown. The typical approach is to assume a uniform, linear, or cubic data distribution within these partially covered intervals. While the linear and cubic method provides a more accurate estimation, we propose the uniform one due to two reasons.
Firstly, applying linear or cubic interpolation entails calculating and recording fixed points, followed by aligning them with linear or cubic curves. However, the integral of these curves might diverge from the corresponding tensor entry, as demonstrated in Figure~\ref{fig:ctn}(b). In contrast, the uniform assumption requires significantly less computation and fewer records on fixed points compared to the linear and cubic alternatives. 
Secondly, the error introduced by fitting a curve to the data distribution is relatively insignificant when compared to the densities of other intervals, as illustrated in example~\ref{ex:density}. Thus, the coefficient can be approximated as proportional to the ratio of the interval covered by the query.

Finally, Equation~\ref{eq:phicon} only consists of multiplications and additions. we proceed to determine the upper bound of multiplications and additions required.
\begin{align}
\begin{split}
    N(\times) & \leq |T_Q| \cdot R + 2 \cdot R \\
    N(+) & \leq  (R - 1) + R \cdot \sum_{T_j \in T_Q} (|\mathbb{T}_j|-1)
\end{split}
\end{align}
When comparing Equation~\ref{eq:phi} with Equation~\ref{eq:phicon}, both involve $R$ iterations. However, the continuous case requires more computations per iteration. Specifically, the number of multiplications increases by $2$ in each iteration because only the intervals at the two ends have coefficients less than $1$, necessitating additional multiplication. In contrast, the intervals in between do not require multiplication in practical implementations. Additionally, for additions, each iteration introduces $\sum_{T_j \in T_Q} (|\mathbb{T}_j|-1)$ more calculations.

We observe a significant increase in the number of additions, but still on the order of $R$. On the other hand, the number of multiplications is only slightly higher than what is observed in the discrete case.

\section{Experiment}
\label{sec:experiment}

In this section, we evaluate \oursys from various perspectives. We first define the general setup of the experiments in Section~\ref{subsec:setup}. Next, we evaluate \oursys on discrete and continuous datasets respectively in Section~\ref{subsec:discrete} and Section~\ref{subsec:continuous}. Moreover, we provide proof that zero-queries are accurately estimated in Section~\ref{subsec:zero}. In Section~\ref{subsec:ablation}, we evaluate the influence of the number of filters and the parameter $R$ in \oursys. Finally, we examine how \oursys behaves when data updates in Section~\ref{subsec:update}. 

\subsection{Experiment Setup}
\label{subsec:setup}

We first describe the general setups across all experiments. The detailed setups are further introduced in each subsection.

\stitle{Datasets.}
\underline{\textit{DMV} dataset}~\cite{misc_dmv} is widely used in many other cardinality estimation works. It records vehicle, snowmobile, and boat registrations in New York. It consists of $11$ columns and $11,591,877$ rows. $10$ out of $11$ attributes are categorical and the remaining one is the date value which can be treated as a continuous attribute. We use both the discretized and the original continuous version of the dataset experiment.

\underline{\textit{Forest} dataset}~\cite{misc_covertype_31} is used as a classification problem involving 7 distinct forest cover types. This dataset consists of a range of attributes, including elevation, aspect, slope, hillshade, soil-type, and more. In our Experiment, we focus on 45 categorical attributes present in the dataset. The covertype dataset comprises a total of 581,011 rows of instances.
    
    \underline{\textit{Poker Hand} dataset}~\cite{misc_poker_hand_158} comprises records of five playing cards randomly drawn from a standard deck of $52$ cards. Its primary objective is to forecast different poker hands. With a total of $11$ attributes and a vast collection of $999,999$ rows.

\stitle{Workloads.}
We generate 12,000 queries, from which 10,000 queries are used for training each query-driven method similar to \cite{pvldb/naru2019, pvldb/YangKLLDCS20} and 2,000 queries are kept for testing all methods. 
The attributes of the table have an independent and identical probability to appear in a query. All the queries have at least two filters as the marginal distributions are too simple.
The operators in each filter predicate vary from $\{>, <, =\}$, and the predicate value is uniformly selected from the corresponding column. 
To fit the reality and further test the accuracy of the models, we only allow equality filters on discrete attributes and inequality filters on continuous attributes.
Furthermore, the queries with zero results are separately evaluated with non-zero queries as they have different difficulties in estimation.

\stitle{Baselines.} We compare \oursys with six state-of-the-art candidate cardinality estimation models as follows:

(1) \mscn~\cite{cidr2019/mscn}. This method encodes the query features with a table set, a join set, and a predicate set using one-hot vectors. Then, it utilizes the multi-set convolutional network to learn a mapping function from the feature set to the predicated cardinality.

(2) \xgb~\cite{pvldb/DuttWNKNC19}. It employs a tree-based ensemble method, called XGBoost \cite{chen2016xgboost}, which encodes a query as a sequence of selection ranges, then learns a mapping function from the query encoding to the predicted cardinality.

(3) \nn~\cite{pvldb/DuttWNKNC19}. This approach is based on fully connected neural networks. It trains a local neural network (NN) model on a fixed join path, and encodes each predicate as four numbers indicating the operations ($<$,$>$,$=$) and the normalized value.

(4) \deepdb~\cite{vldb2020/deepdb}. This method relies on relational sum-product networks (RSPN). It divides a table into row clusters and column clusters recursively. Then it utilizes sum nodes (resp. product nodes) to combine the row clusters (resp. column clusters).

(5) \bayes~\cite{wu2020bayescard}. This method relies on Bayes Network to learn a joint distribution. We adopt a state-of-the-art implementation with probabilistic programming. 

(6) \naru~\cite{pvldb/YangKLLDCS20}. It builds a deep autoregressive model on the samples of the table, then conducts a progressive sampling to make the estimates.

(7) \fctj~\cite{factorjoin}. It combines the classical join-histogram with the learning-based methods to capture attribute distribution efficiently and accurately.

(8) \texttt{UAE}~\cite{wu2021unified}. This method proposes differentiable progressive sampling via the Gumbel-Softmax trick to enable learning from queries. Then it unifies both query and data information using the deep autoregression model.

\stitle{Metrics.}
For non-zero queries, we use q-error to measure their accuracy. The q-error is defined as follows:
\begin{equation*}
    \resizebox{0.95\hsize}{!}{$q = \max \Big( \frac{[estimated \ density]}{true \ density}, \frac{true \ density}{\max(1,[estimated \ density])} \Big)$}
\end{equation*}

where $[\cdot]$ means rounding to nearest integer. We will report $50\%$, $95\%$, and $99\%$ quantiles, but the mean and the max q-errors are excluded. The reason is the results are heavily dependent on the random seeds, and a few extreme samples can dramatically change the mean and the max. Therefore, the median ($50\%$) is a better metric to represent the average accuracy.

For evaluations on zero queries, we calculate the accuracy rate, which is the ratio of queries that are estimated correctly.
$$
accuracy = \frac{\#\{estimated \ density < 0.5 \}}{\#total \ queries}
$$

In addition to measuring errors, we also evaluate the inference latency of the methods. We capture the average execution time of each query, denoted in milliseconds (ms).

\stitle{Environment.}
All experiments were implemented in Python, and performed on a server with a 20-core Intel(R) Xeon(R) 6242R 3.10GHz CPU, and 256GB DDR4 RAM.
For a fair comparison, all the models are estimated on CPU as not all the models support GPU training and inference.

\subsection{Evaluation on Discrete Datasets}
\label{subsec:discrete}

To process the \textit{DMV} dataset, our initial step involves converting it into a discrete version. Since all attributes, except for '$Reg\_Valid\_Date$,' are categorical in nature, we proceed by categorizing the values of '$Reg\_Valid\_Date$' into $14$ distinct categories based on the corresponding year of the data. Subsequently, we evaluate queries on this discretized version of the dataset.
The domain size of the dataset is about $2.25 \times 10^{13}$, with attribute domain sizes ranging from $2$ to $225$. Given this diversity, balancing the attribute domain sizes through thoughtful joins before applying the covering design becomes crucial. we propose joining the attribute '$Record\_Type$' with '$Scofflaw\_Indicator$' and '$Suspension\_Indicator$', as well as joining '$Fuel\_Type$' with '$Revocation\_Indicator$'. As a result, we have an $8$-column dataset, over which we deploy a $C(8, 4, 3)$ covering design comprising $14$ blocks, each varying in length from $4$ to $7$.
To train tensor decompositions effectively on these blocks, our focus is on minimizing the error of reconstructing the tensor, as it directly influences \oursys's q-error. Each block causes distinct challenges due to varying tensor sizes and correlations, so the hyper-parameter $R$ is necessary to be individually determined for each block. It is noteworthy that $R$ need not be the same across all blocks. In practice, some blocks are required to train multiple times to find the appropriate $R$, ensuring minimal reconstruction error. As a result, the values of $R$ vary from $2,000$ to $15,000$.
To construct a query workload, we ensure that each attribute has an independent $50\%$ chance of appearing in a query. Queries that possess only one filter or have a density equal to zero are excluded from the workload. Since the dataset is discrete, we only allow equality to appear in the filters. By implementing these criteria, we generate a set of $2,000$ queries that satisfy the specified conditions.
In terms of computational cost, the training process takes a total of 6,095 seconds, with an average of 435 seconds per block. Additionally, the local storage required for the tensor decomposition is 142MB in total, averaging 10MB per block.

For the \textit{Forest} dataset, the domain size is about $1.23 \times 10^{14}$. While most attributes possess a domain size of $2$, the '$Cover\_Type$' attribute consists of $7$ distinct values. The Domain sizes are fairly evenly distributed across attributes, therefore joining attributes is not necessary. Given its larger attribute count and domain size compared to \textit{DMV}, we consider using more blocks to cover the table. Thus, we employ a $C(45, 20, 4)$ covering design for attributes, yielding a total of $67$ blocks.
In contrast to \textit{DMV}, the sizes of block tensors in \textit{Forest} are relatively smaller, leading to the training of tensor decompositions with lower $R$ values. The range of $R$ spans from $1,000$ to $2,000$.
The workload of \textit{Forest} is generated in a similar way. However, in this case, each attribute is assigned a $10\%$ chance of being selected in a query.
In terms of computational cost, the training process takes a total of 15,402 seconds, with an average of 230 seconds per block. Additionally, the local storage required for the tensor decomposition is 25.1MB in total, averaging 385KB per block.

For the \textit{Poker} dataset, the domain size is about $3.8 \times 10^9$. The domain sizes of attributes vary from $4$ to $13$. The primary objective of this dataset is to predict the attribute '$Poker\_Hand$', while all other attributes are treated as mutually independent. In line with this, all the queries generated for evaluation involve '$Poker\_Hand$', while other attributes are assigned a $50\%$ chance of selection. Similarly, each block is designed to involve '$Poker\_Hand$'. To accomplish this, we employ a $C(10, 5, 3)$ covering design for attributes excluding '$Poker\_Hand$', with the addition of '$Poker\_Hand$' in each block. In total, $17$ blocks are trained, each consisting of $6$ attributes. For the $R$ value, we opt for a consistent choice of $10,000$ across all blocks.
In terms of computational cost, the training process takes a total of 9,685 seconds, with an average of 570 seconds per block. Additionally, the local storage required for the tensor decomposition is 70.4MB in total, averaging 4.1MB per block.

\begin{table}[!t]
    \centering
    {
    \caption{Evaluation on Discritized Dataset.}
    \label{table:eval_discrete}
    \begin{tabular}{| c | c | c | c | c | c |} 
    \cline{1-6}
    Dataset & Method & $50\%$ & $95\%$ & $99\%$ & Latency(ms) \\
    \cline{1-6}
    \multirow{7}{*}{\dmv} & \oursys & \textbf{1.0} & \textbf{2.25} & \textbf{9.0} & \textbf{0.0468} \\ 
    \cline{2-6}
    & \naru & 1.2 & 5.10 & 11.7 & 3.14 \\
    \cline{2-6}
    & \deepdb & 1.53 & 41.9 & 379 & 1.35 \\
    \cline{2-6}
    & \bayes & 1.15 & 12.2 & 141 & 1.10 \\
    \cline{2-6}
    & \mscn & 7.88 & 83.0 & 268 & 0.0911 \\
    \cline{2-6}
    & \xgb & 4.40 & 75.4 & 599 & 0.0142 \\
    \cline{2-6}
    & \nn & 5.07 & 190 & 2710 & \textbf{0.0051} \\
    \cline{2-6}
    & \fctj & 1.27 & 3.17 & 12.49 & 1.8661   \\
    \cline{2-6}
    & \uae & 1.24 & 2.29 & 3.12 & 6.7505   \\
    \cline{1-6}
    \multirow{7}{*}{\covertype} & \oursys & \textbf{1.0} & \textbf{1.002} & \textbf{1.057} & \textbf{0.0453}\\ 
    \cline{2-6}
    & \naru & 1.10 & 1.83 & 3.50 & 4.33 \\
    \cline{2-6}
    & \deepdb & 1.05 & 3.0 & 14.3 & 0.834 \\
    \cline{2-6}
    & \bayes & 1.02 & 1.80 & 5.68 & 0.926\\
    \cline{2-6}
    & \mscn & 2.08 & 16.5 & 29.3 & 1.57 \\
    \cline{2-6}
    & \xgb & 2.63 & 11.1 & 25.0 & 0.0456 \\
    \cline{2-6}
    & \nn & 1.09 & 1.42 & 4.18 & \textbf{0.0079} \\
    \cline{2-6}
    & \fctj & 1.06 & 1.92 & 6.21 & 1.9943   \\
    \cline{2-6}
    & \uae & 1.18 & 1.76 & 2.73 & 7.3752   \\
    \cline{1-6}
    \multirow{7}{*}{\textit{Poker}} & \oursys & \textbf{1.0} & \textbf{2.0} & \textbf{3.0} & \textbf{0.0804}\\ 
    \cline{2-6}
    & \naru & 1.24 & 6.67 & 6.73 & 6.18 \\
    \cline{2-6}
    & \deepdb & 15.0 & 100 & 122 & 0.815 \\
    \cline{2-6}
    & \bayes & 1.28 & 5.99 & 20.1 & 1.27 \\
    \cline{2-6}
    & \mscn & 1.90 & 4.0 & 8.0 & 0.0494 \\
    \cline{2-6}
    & \xgb & 1.71 & 4.12 & 6.59 & \textbf{0.0257} \\
    \cline{2-6}
    & \nn & 1.34 & 2.98 & 4.98 & 0.0309 \\
    \cline{2-6}
    & \fctj & 1.30 & 7.51 & 8.39 & 2.0653   \\
    \cline{2-6}
    & \uae & 1.20 & 4.29 & 5.20 & 8.5548   \\
    \cline{1-6}
    \end{tabular}
    }
    \vspace{-0.5em}    
\end{table}

The experimental results conducted on the \textit{DMV}, \textit{Forest}, and \textit{Poker} datasets are displayed in Table~\ref{table:eval_discrete}. A general observation is that for all the datasets, the \oursys is the most accurate method. Furthermore, \oursys is faster than all other data-driven methods, and only slower than some workload-driven methods.

In terms of accuracy analysis, we first notice that \oursys not only attains a perfect $50\%$ q-error for all the datasets but also ensures $55.05\%$, $66.95\%$, and $53.45\%$ absolute accuracy on the \textit{DMV}, \textit{Forest}, and \textit{Poker} datasets, respectively.
A notable observation is that data-driven approaches, such as \textit{Naru}, \textit{DeepDB}, and \textit{BayesCard}, tend to outperform workload-driven methods, including \textit{MSCN}, \textit{LW-XGB}, and \textit{LW-NN}, in terms of accuracy. This superiority can be attributed to data-driven methods' ability to directly learn data distribution, unlike workload-driven counterparts that only rely on query workload patterns. However, given the random nature of our experiment workloads, workload-driven methods exhibit less effectiveness due to the absence of specific patterns for learning.
\textit{BayesCard} outperforms \textit{Naru} in terms of $50\%$ q-error, but lags behind at $99\%$ q-error. This trend extends to other methods, including \oursys. Such variations may arise due to differences between continuous and discrete datasets. Some methods treat all values as continuous, forcing a smooth curve aligning all discrete values, averaging out errors from extreme cases, and potentially impacting their performance. 
Another instance is the contrasting behavior of \textit{LW-XGB} and \textit{LW-NN}, with the former performing better on the \textit{DMV} dataset while the latter excels on the \textit{Forest} dataset. Such distinctions can be attributed to method sensitivity towards query complexity. Specifically, query difficulty closely correlates with the number of filters, a factor we will further explore in subsequent experiments.

Turning our focus to latency analysis, we observe that \oursys, as a data-driven approach, exhibits latency at least $16$ times quicker than all other data-driven techniques. When compared to workload-driven methods, \oursys remains faster than \textit{MSCN} on \textit{DMV} and \textit{Forest} datasets, yet slightly slower than \textit{LW-XGB} and \textit{LW-NN}. Workload-driven methods generally exhibit superior speed since they can directly process queries without the need for data distribution considerations. Latency variance across methods and datasets can be attributed to many reasons such as model complexity and query difficulty.

\subsection{Evaluation on Continuous Dataset}
\label{subsec:continuous}

While \oursys is specifically designed for discrete data, our objective is not to demonstrate superiority over other methods exclusively for pure continuous datasets. Nonetheless, it remains crucial to showcase \oursys's capability in handling datasets with a minority of continuous attributes. In this experiment, we evaluate \oursys alongside other methods using the original continuous version of the \textit{DMV} dataset.

The continuous version of \oursys differs from its discretized counterpart solely in the '$Reg\_Valid\_Date$' attribute. Corresponding adjustments are made in workload generation. In this case, queries are composed of equalities ${=}$ for discrete attributes and inequalities ${<,>}$ for the '$Reg\_Valid\_Date$' attribute. The covering design and subsequent training process remain identical to the discretized version. However, the estimation process diverges.

\begin{table}[!t]
    \centering
    {
    \caption{Evaluation on Continuous \textit{DMV} Dataset.}
    \label{table:eval_continuous}
    \begin{tabular}{| c | c | c | c | c |} 
    \cline{1-5}
    Method & $50\%$ & $95\%$ & $99\%$ & Latency(ms) \\
    \cline{1-5}
    \oursys & \textbf{1.0} & \textbf{2.0} & 12.0 & \textbf{0.0559} \\ 
    \cline{1-5}
    \naru & 1.27 & 5.0 & \textbf{7.63} & 4.425\\
    \cline{1-5}
    \deepdb & 1.43 & 33.9 & 196 & 1.2 \\
    \cline{1-5}
    \bayes & 1.12 & 13.3 & 140 & 2.56 \\
    \cline{1-5}
    \mscn & 6.39 & 102 & 351 & 0.08 \\
    \cline{1-5}
    \xgb & 4.29 & 86.0 & 1130 & \textbf{0.031} \\
    \cline{1-5}
    \nn & 4.40  & 209 & 2440 & 0.082 \\
    \cline{1-5}
    \fctj & 1.272 & 7.7161 & 38.0 & 2.574   \\
    \cline{1-5}
    \uae & 1.27 & 3.52 & 7.81 & 8.558   \\
    \cline{1-5}
    \end{tabular}
    }
    \vspace{-0.5em}
\end{table}

The experimental results are displayed in Table~\ref{table:eval_continuous}. Results show that \oursys performs as the top method for $50\%$ and $95\%$ q-errors, slightly trailing \textit{Naru} for $99\%$ q-error. Impressively, \oursys ensures the absolute estimation accuracy of $53.25\%$ of queries. Concerning inference latency, \oursys ranks as the second fastest method, trailing only \textit{LW-XGB}. In general, most methods demonstrate similar behavior to the discretized \textit{DMV} experiment. Notably, minor differences arise not only from the contrast between continuous and discrete datasets but also from the influence of randomly generated queries. However, \textit{LW-NN} significantly diverges between the two datasets, with markedly slower inference latency in the continuous version compared to its discrete counterpart.

\begin{table}[!t]
    \centering
    {
    \caption{Evaluation of \oursys on Continuous DMV with Varying Number of Bins.}
    \label{table:CoDe_on_DMV_vbins}
    \begin{tabular}{| c | c | c | c | c |} 
    \hline
    \# bins & $50\%$ & $95\%$ & $99\%$ & Latency(ms) \\
    \hline
    9  & 1.001 & 23.9 & 307 & 0.0713\\
    \hline
    14  & 1.0 & 2.0 & 12.0 & 0.0559\\
    \hline
    20  & 1.0 & 3.25 & 22.53 & 0.0792\\
    \hline
    \end{tabular}}
    \vspace{-0.5em}    
\end{table}

By varying the number of bins, we observe from Table~\ref{table:CoDe_on_DMV_vbins} that the model becomes less accurate when the number of bins is too small. This inaccuracy primarily stems from the large bin sizes. However, other factors also impact accuracy. For example, an excessively large number of bins increases the domain size, making the model harder to train. Additionally, the model's inherent perturbation during training further influences its accuracy.

\subsection{Evaluation on Zero-queries}
\label{subsec:zero}

\begin{table}[!t]
    \centering
    {
    \caption{Accuracy on Zero-queries.}
    \label{table:eval_zero}
    \begin{tabular}{| c | c | c | c |} 
    \cline{1-4}
    Method & \dmv & \covertype & \textit{Poker} \\
    \cline{1-4}
    \oursys & \textbf{99.1\%} & \textbf{96.5\%} & \textbf{96.1\%} \\
    \cline{1-4}
    \naru & 29.5\% & 24\% & 26.2\% \\
    \cline{1-4}
    \deepdb & 48.7\% & 42.6\% & 44.4\% \\
    \cline{1-4}
    \bayes & 25.1\% & 30.1\% & 28.7\% \\
    \cline{1-4}
    \mscn & 5.1\% & 6.7\% & 6.3\% \\
    \cline{1-4}
    \xgb & 4.7\% & 4.9\% & 4.9\% \\
    \cline{1-4}
    \nn & 4.2\% & 5.1\% & 4.6\% \\
    \cline{1-4}
    \uae & 23.9\% & 12.8\% & 15.6\% \\
    \cline{1-4}
    \fctj & 24.7\% & 28.4\% & 27.6\% \\
    \cline{1-4}
    \end{tabular}
    }
    \vspace{-0.5em}    
\end{table}

As illustrated in Figure~\ref{fig:sampling}, queries are highly likely to yield zero instances when the number of filters is high enough. This situation prompts a significant challenge for existing cardinality estimation methods due to they lack a consideration for these zero-queries. Ensuring the proper functioning of \oursys for zero-queries becomes a pivotal consideration. In this experiment, we generate training queries similar to the previous trials conducted on the \textit{DMV}, \textit{Forest}, and \textit{Poker} datasets. The key difference is that we do not eliminate queries with zero cardinalities (we assign a true cardinality of $1e^{-3}$ to queries with zero cardinalities because true cardinalities should be greater than zero to calculate q-error). For generating testing queries, we specifically retain the zero-queries from the generated queries. We evaluate all methods based on their accuracy in correctly estimating the ratio of queries, with estimated densities below $0.5$ considered as accurate.

The outcomes of this experiment are summarized in Table~\ref{table:eval_zero}. Impressively, \oursys significantly outperforms all other methods, successfully estimating nearly all zero-queries. The second-best performer is \textit{DeepDB}, which manages to correctly estimate less than half of the queries. Conversely, workload-driven methods struggle to estimate any zero-queries correctly. This reason can be attributed to their training on workload dominated by non-zero cardinality queries. This type of training is necessary for workload-driven methods to accurately estimate most queries, but it can introduce a bias towards producing non-zero answers. This exposes a significant limitation of workload-driven methods, which is their heavy reliance on historical workloads rather than actual data, making them inadequate for handling unfamiliar queries.

\subsection{Ablation Study}
\label{subsec:ablation}

In this section, our exploration delves into exploration of the impact of pivotal parameters: the number of filters within the queries and the parameter $R$, $k$, and $t$.

\begin{table}[!t]
    \centering
    {
    \caption{Evaluation of \oursys against $\#$filters.}
    \label{table:eval_filter}
    \begin{tabular}{| c | c | c | c | c | c |} 
    \cline{1-6}
    Dataset & $\#$filters & $50\%$ & $95\%$ & $99\%$ & Latency(ms) \\
    \cline{1-6}
    \multirow{4}{*}{\dmv} & 2 & 1.0 & 1.69 & 4.0 & 0.0469\\ 
    \cline{2-6}
    & 3 & 1.0 & 2.0 & 4.0 & 0.0472\\
    \cline{2-6}
    & 4 & 1.0 & 3.0 & 13.01 & 0.0632\\
    \cline{2-6}
    & 5 & 1.0 & 5.0 & 22.02 & 0.0634\\
    \cline{1-6}
    \multirow{4}{*}{\covertype} & 2 & 1.0 & 1.0003 & 1.008 & 0.0396\\ 
    \cline{2-6}
    & 4 & 1.0 & 1.01 & 1.21 & 0.0436\\
    \cline{2-6}
    & 6 & 1.0 & 1.09 & 2.10 & 0.0737\\
    \cline{2-6}
    & 8 & 1.001 & 1.33 & 3.87 & 0.0862\\
    \cline{1-6}
    \multirow{4}{*}{\textit{Poker}} & 2 & 1.0 & 1.03 & 1.13 & 0.0431\\ 
    \cline{2-6}
    & 3 & 1.0 & 1.16 & 1.5 & 0.0465\\
    \cline{2-6}
    & 4 & 1.0 & 1.10 & 2.0 & 0.0509\\
    \cline{2-6}
    & 5 & 1.0008 & 2.0 & 3.0 & 0.0809\\
    \cline{1-6}
    \end{tabular}
    }
    \vspace{-0.5em}    
\end{table}

To begin, we investigate the impact of the number of filters in queries on the performance of \oursys. This experiment employs the models trained in Section~\ref{subsec:discrete}. We proceed to generate multiple workloads for each dataset, maintaining a consistent pattern while varying the number of filters in each workload. For the \textit{DMV} and \textit{Poker} datasets, filter counts range from $2$ to $5$. while for the \textit{Forest} dataset, filter counts span from $2$ to $8$.

The outcomes are presented in Table~\ref{table:eval_filter}. A noteworthy observation is that as the number of filters increases for all the datasets, both the errors and latency increase. A higher number of filters corresponds to larger domain sizes, making accurate query estimation more challenging. Likewise, a greater number of filters entails increased multiplicative operations, consequently lengthening inference time. Notably,  the median q-errors remain relatively stable at precisely $1$, except for the cases of $8$ filters on the \textit{Forest} dataset and $5$ filters on the \textit{Poker} dataset. This indicates that \oursys proficiently resolves most queries, even when the number of filters is substantial. Notably, a latency surge arises between the $3$-filter and $4$-filter cases for the \textit{DMV} dataset. This is due to the fact that while $3$-filter queries can be fully covered by a single block, $4$-filter queries might require two blocks for comprehensive resolution. A similar trend happens between the $4$-filter and $6$-filter scenarios for the \textit{Forest} dataset, and the $4$-filter and $5$-filter scenarios for the \textit{Poker} dataset.

In a subsequent exploration, we delve into the impact of the parameter $R$ on \oursys performance. This experiment employs the workload generated in Section~\ref{subsec:discrete}, and the models are trained with varying choices of $R$ values. Notably, all blocks share the same $R$ value for each setting, with training executed once irrespective of errors. For the \textit{DMV} and \textit{Poker} datasets, $R$ values span from $2,000$ to $10,000$, while for the \textit{Forest} dataset, the range is $1,000$ to $5,000$.

\begin{table}[!t]
    \centering
    {
    \caption{Evaluation of \oursys against $R$.}
    \label{table:eval_R}
    \begin{tabular}{| c | c | c | c | c | c |} 
    \cline{1-6}
    Dataset & $R$ & $50\%$ & $95\%$ & $99\%$ & Latency(ms) \\
    \cline{1-6}
    \multirow{3}{*}{\dmv} & 2000 & 1.00007 & 20.1 & 940 & 0.0393\\ 
    \cline{2-6}
    & 5000 & 1.00002 & 7.52 & 96.0 & 0.0479\\
    \cline{2-6}
    & 10000 & 1.0 & 5.0 & 35.0 & 0.0554\\
    \cline{1-6}
    \multirow{3}{*}{\covertype} & 1000 & 1.0 & 1.003 & 1.06 & 0.0449\\ 
    \cline{2-6}
    & 2000 & 1.0 & 1.014 & 1.125 & 0.0471\\
    \cline{2-6}
    & 5000 & 1.0 & 1.025 & 1.195 & 0.0537\\
    \cline{1-6}
    \multirow{3}{*}{\textit{Poker}} & 2000 & 1.06 & 2.0 & 4.0 & 0.0557\\ 
    \cline{2-6}
    & 5000 & 1.02 & 2.0 & 3.0 & 0.0653\\
    \cline{2-6}
    & 10000 & 1.0 & 2.0 & 3.0 & 0.0804\\
    \cline{1-6}
    \end{tabular}
    }
    \vspace{-0.5em}    
\end{table}

The outcomes are presented in Table~\ref{table:eval_R}. An initial observation reveals that an increase in $R$ leads to a subsequent rise in inference latency across all the datasets. This correlation can be attributed to the fact that both the number of additions and multiplications are proportional to $R$. However, the relationship between latency and $R$ is not strictly proportional, as block selection also requires inference time. Furthermore, concerning errors, an interesting trend emerges: while the errors on the \textit{DMV} and \textit{Poker} datasets decrease as $R$ grows, the opposite trend is observed for the \textit{Forest} dataset. This behavior arises due to the fact that a larger $R$ is not necessarily better. Excessive $R$ values can lead to tensor decomposition reconstruction errors oscillating around zero, consequently resulting in higher q-errors in estimation.

\begin{table}[!t]
    \centering
    {
    \caption{Evaluation of \oursys on DMV with varying $k$ and $t$.}
    \label{table:eval_kt}
    \begin{tabular}{| c | c | c | c | c |} 
    \hline
    Covering Design & $50\%$ & $95\%$ & $99\%$ & Latency(ms) \\
    \hline
    $C(8,4,3)$  & \textbf{1.0} & \textbf{2.25} & \textbf{9.0} & \textbf{0.0468}\\
    \hline
    $C(11,4,3)$  & 1.0008 & 29.3 & 291 & 0.0556\\
    \hline
    $C(11,4,2)$  & 1.024 & 43.0 & 378 & 0.0603\\
    \hline
    $k \ge 5$  & \multicolumn{4}{c|}{Failed to compute}\\
    \hline
    \end{tabular}}
    \vspace{-0.5em}    
\end{table}

We evaluate the effects of parameters $k$ and $t$. In Table~\ref{table:eval_kt}, $C(8,4,3)$ represents the recommended method from Section~\ref{subsec:discrete}, which utilizes covering design after join operations. Theoretically, larger values for both $k$ and $t$ are preferable. Our experiments specifically compare cases where $k=4$ versus $k \ge 5$. Notably, the situation without any covering design occurs when $k=v$.
However, when $k=5$, tensor decomposition becomes infeasible due to excessive domain size. For instance, the subspace formed by the five largest attributes in the DMV dataset has a domain size of $5.2 \times 10^9$, requiring approximately 20GB of storage space alone - making tensor decomposition computationally prohibitive. This clearly demonstrates the critical importance of covering design.
Comparing $C(11,4,3)$ with $C(8,4,3)$, the former shows inferior performance. While both share the same $t$ value, $C(8,4,3)$ requires fewer blocks, allowing for more targeted adjustments or retraining. Among all configurations, $C(11,4,2)$ performs the worst due to its minimal $t$ value, which results in many queries not being covered by a single block.

\subsection{Data Update}
\label{subsec:update}

\begin{table}[!t]
    \centering
    {
    \caption{Evaluation of Dataset Update.}
    \label{table:update}
    \begin{tabular}{| c | c | c | c | c |} 
    \cline{1-5}
    Dataset & $50\%$ & $95\%$ & $99\%$ & Update Time(ms) \\
    \cline{1-5}
    \textit{DMV} & 1.011 & 10.2 & 96.0 & 91.0 \\ 
    \cline{1-5}
    \textit{Forest} & 1.0006 & 1.011 & 1.058 & 52.9\\
    \cline{1-5}
    \textit{Poker} & 1.03 & 2.0 & 3.0 & 71.2 \\
    \cline{1-5}
    \end{tabular}
    }
    \vspace{-0.5em}    
\end{table}

When addressing data updates, we explore two distinct methods: updating only the weights and performing a full update. To investigate this, we initially sample $95\%$ of the data from each dataset, simulating the dataset's state before the update. Subsequently, \textit{CoDe} models are trained on these sampled datasets. The weights are then updated by scaling them roughly by a factor of $\frac{1}{0.95}$. The updated model is evaluated using the workload generated in Section~\ref{subsec:discrete}, and the outcomes are presented in Table~\ref{table:update}.

To evaluate the significance of these results, we compare them against the optimum outcomes detailed in Table~\ref{table:eval_discrete}. The findings indicate that q-errors for the \textit{Forest} and \textit{Poker} datasets are slightly worse than the optimum results. Notably, the $95\%$ and $99\%$ q-errors on the \textit{DMV} dataset exceed the expected performance. This can potentially be attributed to the algorithmic perturbations introduced during the update process. Nonetheless, most of these updated results outperform those of all other methods. This implies that in many scenarios, updating only the weights is sufficiently effective, even with a $5\%$ difference in the dataset. Moreover, the weight update process takes less than $0.1$ seconds across all datasets, demonstrating its viability for daily applications. However, it's important to note that this approach no longer guarantees that most queries will be absolutely accurate. Therefore, for scenarios demanding optimal results, full retraining is recommended.

\section{Related Work}
\label{sec:related}

\subsection{Cardinality Estimation.} 
Cardinality estimation is one of the most important components in database optimizer. 
Traditional methods mainly contain histograms and samplings, which with the advantages of small overhead and strong applicability. 
\textit{Histogram}~\cite{selinger1979access} assumes that all attributes are independent and is implemented in DBMS like Postgres~\cite{misc_postgre}. Additionally, multi-dimensional histogram~\cite{deshpande2001independence, gunopulos2000approximating, poosala1997selectivity, wang2003multi,zhang2024htap} variations have been proposed to model distributions more comprehensively.
\textit{Sampling}~\cite{leis2017cardinality, zhao2018random} samples a portion of the dataset for estimation, this approach is used in DBMS such as MySQL~\cite{misc_mysql}. 

Learned cardinality estimation models~\cite{sun2021learned,zhang2023autoce,zhang2024pace,pvldb/naru2019,vldb2020/deepdb,wu2021unified,aytimur2024lplm,park2020quicksel,li2023alece} are based on machine learning techniques. Compared to traditional methods, they are more accurate, but less applicable due to higher overhead and limited support for query types.
In addition to the models outlined in Section~\ref{sec:experiment}, there are several other representative methods:
(1) \textit{Bayesian Networks}~\cite{chow1968approximating, getoor2001selectivity, tzoumas2011lightweight, halford2019approach} (BN). BN-based methods employ a directed acyclic graph to formulate conditional probabilities for dataset attributes. Attributes under the same parent node are assumed to be conditionally independent.
(2) \textit{Deep Auto-Regression}~\cite{hasan2019multi, pvldb/YangKLLDCS20} (DAR) models employ the chain rule for joint distribution formulation. Conditional probabilities are modeled using a deep neural network (DNN).
(3) \textit{FLAT}~\cite{pvldb/FLAT21} is a model based on the factorized-split-sum-product networks (FSPN~\cite{wu2020fspn}).
(4) \textit{ASM}~\cite{kim2024asm} is a new learned cardinality estimator that use autoregressive model, sampling, and multi-dimensional statistics merging for cardinality estimation.

\subsection{Tensor Decomposition.}

Tensor decomposition is a mathematical technique used to break down multi-dimensional arrays (tensors) into simpler, interpretable components. It extends the concept of matrix factorization to higher dimensions, enabling pattern extraction, dimensionality reduction, and efficient computation for complex datasets.
(1) \textit{CP (CANDECOMP/PARAFAC)}~\cite{kolda2009tensor}: Decomposes a tensor into a sum of rank-one tensors. It is simple and unique under mild conditions but lacks flexibility due to its fixed rank structure.
(2) \textit{PARAFAC2}~\cite{kiers1999parafac2, bro1999parafac2}: A variant of PARAFAC designed for tensors with uneven sizes.
(3) \textit{Tucker decomposition}~\cite{tucker1966some}Breaks down a tensor into a core tensor and multiple factor matrices. It is more flexible than CP (allowing different ranks per mode) but is generally non-unique. Notably, CP decomposition can be viewed as a special case of Tucker decomposition where the core tensor is diagonal.
(4)\textit{Tensor train decomposition (TT)}~\cite{oseledets2011tensor}: Represents a tensor as a sequence of matrices, where each entry of the tensor is computed as the product of these matrices. 
(5)\textit{Tensor ring decomposition}~\cite{zhao2016tensor}: An advanced form of TT where the decomposed tensors form a cyclic ring structure. It offers greater flexibility and often achieves better compression compared to TT.

\subsection{Covering Design.}
A universal solution for all covering designs $C(v,k,t)$ does not exist~\cite{gordon1995new, colbourn2010crc, dinitz1992contemporary}. While the $t=2$ scenario is frequently analyzed~\cite{abel2007pair, bluskov2000infinite, greig2006covering, abel2007pair2}, the complexity increases considerably for $t>2$~\cite{bertolo2004upper, bluskov1998new}. Moreover, research has delved into establishing lower bounds for covering designs~\cite{applegate2003asymmetric, furedi1989projective, greig2006covering, horsley2017generalising, horsley2018new}.
Besides covering design, several other mathematical concepts share similarities with it: 
(1)\textit{Lotto Designs}: A generalization of covering designs. A $(v,k,t,m)$-lotto design ensures that for any $t$-element subset (the "draw"), at least one $k$-element block (a "ticket") intersects it in at least $m$ elements.
(2)\textit{Steiner Systems}: These require every $t$-subset to appear in exactly one block.
(3)\textit{Packing Designs}: These limit $t$-subsets to appearing in at most one block, optimizing non-overlapping coverage.

\section{Conclusion and Future Work}
\label{sec:conclusion}

This paper introduces a novel cardinality estimation method named \oursys. By embracing the covering design strategy, we capture dataset distribution through multiple models rather than a single one. Each model is built using the tensor decomposition technique. \oursys demonstrates outstanding accuracy and speed.

There are two most interesting problems that deserve further study. Firstly, exploring more flexible and efficient table covering methods could enhance the approach. For example, similar to~\cite{sagerag,zhang2025spargeattn, zhang2024sageattention2, zhang2025sageattention,zhang2025sageattention2++}, employing the covering method on GPUs in parallel could further reduce the latency of \oursys, making it more likely to be used in production scenarios.
Secondly, extending \oursys to scenarios like fully continuous datasets, multi-table setups, and cardinality estimation tasks in federated learning presents promising directions.

\begin{acknowledgment}
This paper was supported by National Key R\&D Program of China (2023YFB4503600),
NSF of China (62525202, 62232009), Shenzhen Project (CJGJZD20230724093403007), Zhongguancun
Lab, and Beijing National Research Center for Information Science and Technology
(BNRist). Jintao Zhang and Guoliang Li are corresponding authors.
\end{acknowledgment}

\newpage
\bibliographystyle{abbrv}
\bibliography{bibliography}

\vspace{-3em}
\begin{IEEEbiography}[{\includegraphics[width=1in,height=1in,clip,keepaspectratio]{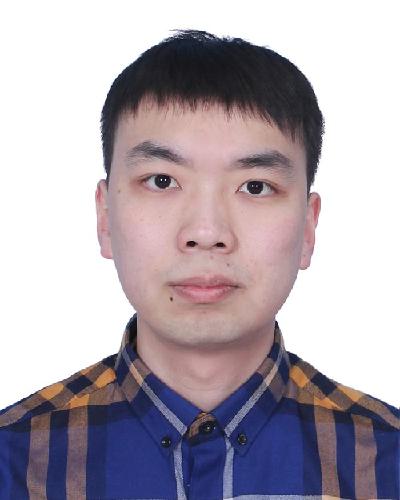}}]{Yaoyu Zhu}
received the MSci degree in mathematics from Imperial College London. He is currently working toward the PhD degree in the Department of Computer Science, Tsinghua University, Beijing, China. His research interests include machine learning and query processing for database.
\end{IEEEbiography}

\vspace{-3em}

\begin{IEEEbiography}[{\includegraphics[width=1in,height=1in,clip,keepaspectratio]{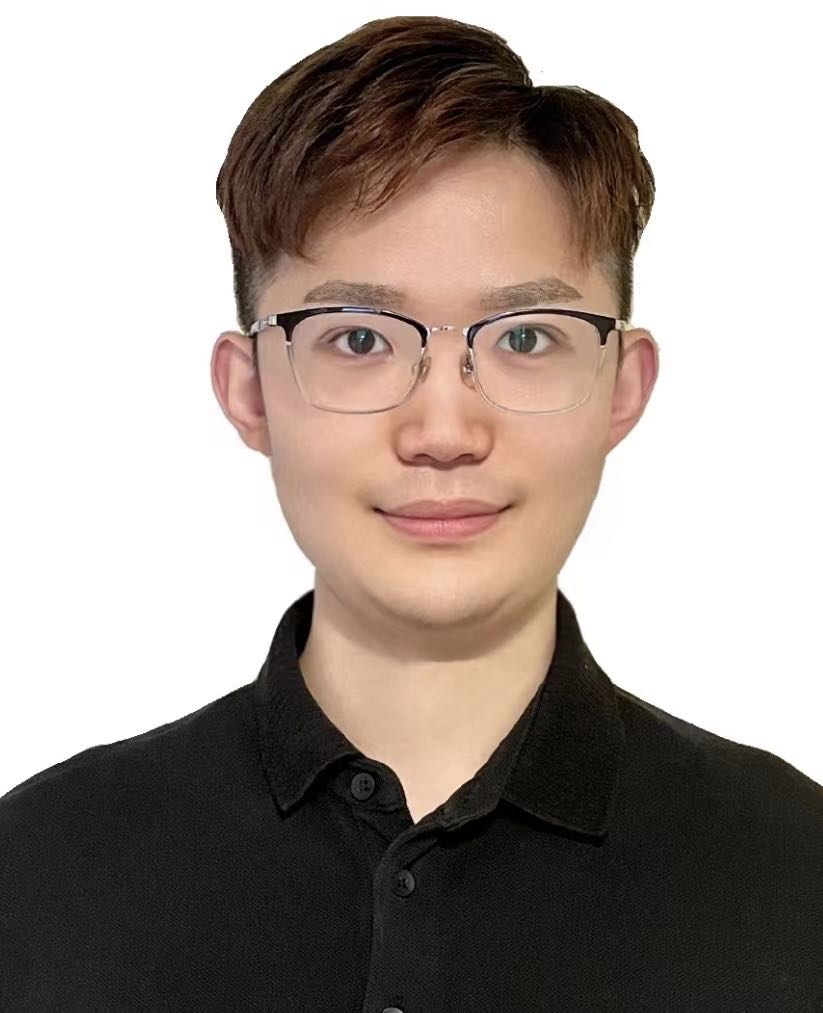}}]{Jintao Zhang}
is working toward his PhD degree in the Department of Computer Science, Tsinghua University, Beijing, China. His research interests include query processing for database.
\end{IEEEbiography}

\vspace{-3em}
\begin{IEEEbiography}[{\includegraphics[width=1in,height=1in,clip,keepaspectratio]{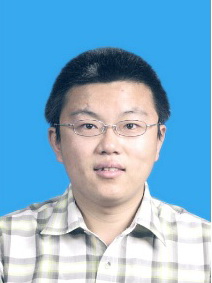}}]{Guoliang Li}
is currently working as a professor in the Department of Computer Science, Tsinghua University, Beijing, China. He received his PhD degree in Computer Science from Tsinghua University, Beijing, China in 2009. His research interests mainly include database systems, data cleaning and integration, and AI$\&$DB co-optimization.
\end{IEEEbiography}

\vspace{-3em}

\begin{IEEEbiography}[{\includegraphics[width=1in,height=1in,clip,keepaspectratio]{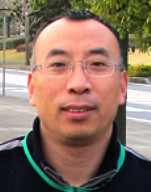}}]{Jianhua Feng}
received the BS, MS, and the PhD degrees in computer science from Tsinghua University. He is currently working as a professor in the Department of Computer Science, Tsinghua University. His main research interests include large-scale data management.
\end{IEEEbiography}

\vfill

\end{document}